\pdfoutput=1
\documentclass{article}

\usepackage{arxiv}

\usepackage[utf8]{inputenc} %
\usepackage[T1]{fontenc}    %
\usepackage{hyperref}       %
\usepackage{url}            %
\usepackage{booktabs}       %
\usepackage{amsfonts}       %
\usepackage{nicefrac}       %
\usepackage{microtype}      %
\usepackage{graphicx}
\usepackage{doi}
\usepackage[font=small]{caption}
\usepackage{xcolor}
\usepackage{subcaption}
\usepackage{float}

\usepackage{amsmath}
\usepackage{amsthm}
\usepackage{amssymb}
\newtheorem{definition}{Definition}
\newtheorem{theorem}{Theorem}

\newtheorem{proposition}{Proposition}

\usepackage[linesnumbered,ruled,vlined]{algorithm2e}
\usepackage{subfiles} 
\usepackage{cite}

\DeclareMathOperator*{\argmax}{argmax}

\title{Measuring Privacy Loss in Distributed Spatio-Temporal Data}

\author{Tatsuki Koga\\
    University of California, San Diego\\
    \texttt{tkoga@ucsd.edu}\\
    \And
    Casey Meehan\\
    Tumult Labs\\
    \texttt{Caseymacmeehan@gmail.com}\\
    \And
    Kamalika Chaudhuri\\
    University of California, San Diego\\
    \texttt{kamalika@cs.ucsd.edu}\\
}

\date{}

\renewcommand{\shorttitle}{Measuring Privacy Loss in Distributed Spatio-Temporal Data}

\begin{document}
\maketitle
\begin{abstract}
Statistics about traffic flow and people's movement gathered from multiple geographical locations in a distributed manner are the driving force powering many applications, such as traffic prediction, demand prediction, and restaurant occupancy reports.
However, these statistics are often based on sensitive location data of people, and hence privacy has to be preserved while releasing them. The standard way to do this is via differential privacy, which guarantees a form of rigorous, worst-case, person-level privacy. In this work, motivated by several counter-intuitive features of differential privacy in distributed location applications, we propose an alternative privacy loss against location reconstruction attacks by an informed adversary.
Our experiments on real and synthetic data demonstrate that our privacy loss better reflects our intuitions on individual privacy violation in the distributed spatio-temporal setting.
\end{abstract}

\section{Introduction}
Statistics about occupancy and people's movement gathered across multiple locations in a distributed manner are the driving force behind a number of everyday applications, such as traffic prediction, demand prediction, and restaurant occupancy reports. However, these statistics are based on individuals’ location data, which is itself sensitive and can be used to infer other sensitive information such as where people live and work. Therefore, it is crucial to understand how people's privacy may be compromised while publishing such statistics, so that this can be done in a more privacy-aware manner.

The standard way to measure privacy in this setting is to use Differential Privacy (DP), which ensures that the participation of a single person in an {\em{entire dataset}} does not change the probability of any outcome by much. This is obtained by adding, to each statistic, noise that is enough to obscure the participation of one person, and guarantee the right amount of privacy. In a distributed spatio-temporal setting, the entire dataset is everyone all over the world, which leads to overly pessimistic privacy accounting. 

To understand this better, let us take a closer look at the kind of adversaries that differential privacy would protect against. Suppose we have an adversary with very specific knowledge about Alice: they know that if Alice is not in NYC, then she is in a specific location in Paris. Any occupancy statistic, released with DP, will provide Alice protection against this adversary---wherever in the world she happens to be. This is excellent from the privacy stand-point, but sometimes leads to pessimistic accounting of loss, and counter-intuitive effects: even if we never have a camera in Paris, and Alice never visits NYC, she still suffers a privacy loss every time a distant camera switches on in NYC.

The example above shows that if we want a tighter accounting of privacy loss, then we need to look at a more constrained of adversaries. A way to measure privacy loss, alternative to differential privacy, is to directly measure privacy loss of reconstruction attacks against an adversary who only knows the prior distribution of the target individual’s locations. \cite{guo_analyzing_2022} propose such a framework for non-temporal discrete data. The framework successfully overcomes the aforementioned issue of DP by introducing the individual prior distribution. However, their framework is not practically applicable to spatio-temporal data because it only measures the privacy loss of reconstructions of all locations at all time steps, and the privacy loss computation time scales exponentially with the number of time steps, $T$, without any assumption.

To this end, we first formally set up an adversary, who is similar to the one in \cite{guo_analyzing_2022}, and define privacy loss of an individual which is suitable in the spatial-temporal setting. In particular, we suppose the adversary knows the prior distribution of the target individual's locations, which are assumed to form the Markov chain. We then use the relaxed definition of the adversary's success---we say the adversary succeeds when they reconstruct some fractions of the location trace. We finally measure the privacy loss by the probability of the adversary's success.

Next, we provide a way to estimate our privacy loss against specific adversarial estimators, e.g., maximum a posteriori (MAP), and upper-bound our loss for any estimator. We show that both estimating and upper-bounding our privacy loss can be done in polynomial time in $T$ under the Markov assumption on the individual's locations, and we provide algorithms for this estimation.

Finally, our experiments on real datasets demonstrate that our privacy loss reflects our intuitions on privacy loss in the distributed spatio-temporal setting. In particular, we show that the privacy loss is higher for individuals who are more likely to be at the place where occupancy statistics are released. The experiments further indicate that individuals who move slowly, i.e., have less uncertainty about where they are at the next time step, generally have higher privacy loss. 
We additionally observe that our privacy loss upper bound gets tighter when the adversary has relatively less information about the target individual.

\subsection{Related Work}
Previous literature has studied location privacy in the context of differential privacy or its variants. 
In particular, there exist several papers investigating the privacy-preserving release of individual location traces.
\cite{andres_geo-indistinguishability_2013} introduces geo-indistinguishability which protects an individual’s location within a certain radius and provides mechanisms to achieve the privacy definition.
\cite{xiao_protecting_2015-1} proposes $\delta$-location set to account for the temporal correlation of location traces.
Apart from DP, \cite{shokri_quantifying_2011} empirically examines several attacks on released location traces. \cite{cao_priste_2019, meehan_location_2021-1} propose inferential privacy definitions, which instantiate Pufferfish privacy~\cite{kifer_pufferfish_2014}, for individual location traces.
Although it is conceptually possible to aggregate all individuals’ location traces, published in a privacy-preserving manner, to release occupancy statistics, it requires everyone to publish their location every time such a statistic is released anywhere on earth. This is not realistic at all due to excessive computational costs. Furthermore, if we consider the case of releasing statistics gathered by physical sensors, it is much more unrealistic—physical sensors can only communicate with nearby people.
There also are some papers that aim to publish privacy-preserving aggregate statistics of spatio-temporal data.
\cite{fang_differential_2014} and \cite{duchi_distance-based_2013} incorporate the distance metric into the neighboring relation of DP. However, the privacy loss is still uniform across individuals; thus, getting back to the previous example, privacy loss for Alice, who is in Paris, is as large as the one for those in NYC when occupancy statistics in NYC are released.
\cite{rafiei_quantifying_2022} studies the potential privacy loss due to temporal correlation when event-level DP is satisfied.

Another line of related work is quantifying privacy loss against reconstruction attacks \cite{bhowmick_protection_2019,nasr_adversary_2021-1,guo_bounding_2022,balle_reconstructing_2022,guo_analyzing_2022,humphries_investigating_2023}. This line of work has emerged from the attempts to understand the fine-grained privacy loss of releasing machine learning models trained with DP. Our work is built on top of this idea of quantifying privacy loss to investigate more realistic privacy loss in the spatio-temporal setting. In addition, while the previous work focuses on relating the privacy loss with DP, our work examines the privacy loss when there are no additive noise, i.e., DP is not satisfied at all.

\section{Preliminaries}
\subsection{Problem Setup}

Suppose there are $M$ discrete locations where sensors may be placed and $N$ people who can be at any one of these locations at each of $T$ time steps. 
We use the notation $X^i_t$ to denote the $i$-th individual's location, and $D=\{ X^{i}_{t}: i = 1, \ldots, N \land t = 1, \ldots, T\}$ to denote a dataset, which is a set of $N$ people's locations over time.
At each time $t \in \{1,\ldots, T\}$, we assume that a sensor at a location $c_t \in \{1, \ldots, M \}$ switches on, and the sensor records the count of people at that location, $\mathrm{count}(D_t,c_t)$, where $D_t = \{ X^{i}_{t}: i = 1, \ldots, N\}$ is the dataset at time $t$.
This count, or an approximate version of it, is then published, and seen by an adversary.

Based on the published counts across time, the adversary aims to identify the target person's locations.
Our goal is to measure the individual privacy loss in this setting.

\subsection{Preliminaries on Privacy}
\subsubsection{Differential Privacy}
The standard way to measure privacy in these kinds of settings is differential privacy, a strong cryptographically-motivated definition of individual-level privacy. Differential privacy guarantees that the participation of a single person in a dataset does not the probability of any outcome by much. Formally, it is defined as follows.

\begin{definition}[$(\epsilon,\delta)$-Differential Privacy~\cite{dwork_our_2006-1}]
A randomized algorithm $\mathcal{M}$ satisfies $(\epsilon,\delta)$-differential privacy if for any two datasets $D,D^{\prime}$ that differ in the value of a single individual, and for any $S\subseteq \mathrm{range}(\mathcal{M})$,
\begin{align*}
    \Pr [\mathcal{M}(D) \in S] \leq \exp (\epsilon) \Pr [\mathcal{M}(D^\prime) \in S]+\delta.
\end{align*}
\end{definition}

In our setting, recall a dataset $D$ is a set of $N$ people's locations over time, i.e., $D = \{ X^{i}_{t}: i = 1, \ldots, N \land t = 1, \ldots, T\}$. When one person changes, their entire set of locations from $t=1, \ldots, T$ changes. Observe that here $\epsilon$ is a parameter which measures privacy loss---lower $\epsilon$ means better privacy. 

The standard differentially private way to address our problem is to add the Gaussian noise with standard deviation $\sigma$ to each count at each location.
In particular, at each time $t$, we release $\mathrm{count}(D_t,c_t) + \xi_t$,
where $\xi_t \sim \mathcal{N}(0,\sigma^2)$ is the added Gaussian noise at time $t$.
Standard analysis~\cite{dwork_algorithmic_2014} shows that this leads to a privacy loss of
$\epsilon = \frac{\sqrt{2\ln(1.25/\delta)\cdot T}}{\sigma}$ for a fixed $\delta$. 

The main pitfall of measuring the individual privacy loss with DP is that \emph{every} person has the same privacy loss. 
This is counter-intuitive for our setting. The following simple example shows how the uniform privacy loss does not reflect the actual privacy loss and what we expect for privacy loss in our setting.

\paragraph{Example 1.}
\begin{figure}[t]
    \centering
    \includegraphics[width=\linewidth]{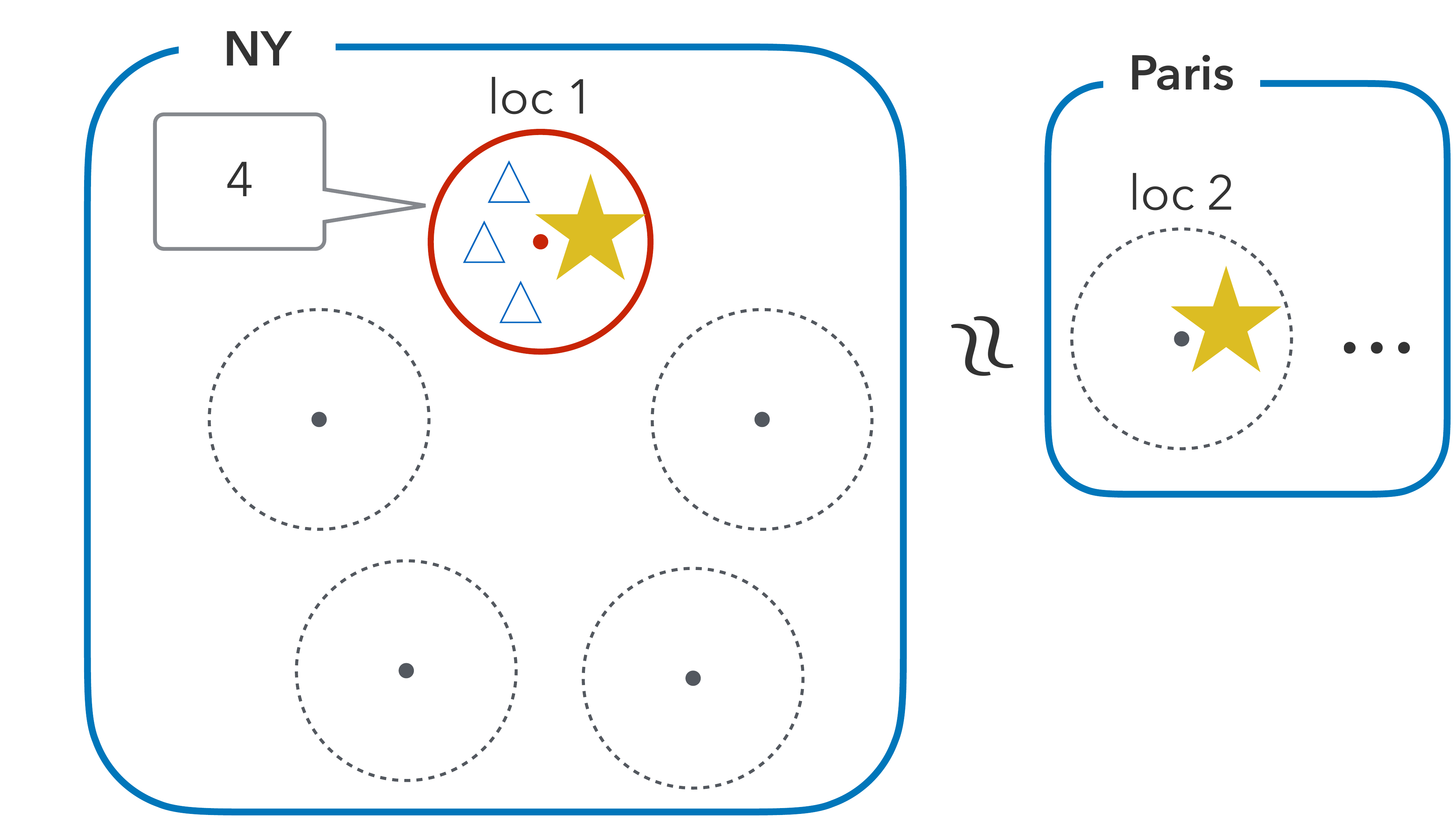}
    \caption{Our distributed spatial counting problem at a fixed time step. Each circle represents a location: a solid red circle indicates the count is published from the location, and dotted black circles indicate it is not. Blue triangles and yellow stars represent people. Since there are three triangles and one star within the red circle, the published count at this time is 4. }
    \label{fig:example1}
\end{figure}
Consider the spatial counting problem in Figure~\ref{fig:example1} where the number of people within the red solid circle is published. 
A person in NY can be at the solid red circle (loc 1) with considerable probability. Therefore, it is reasonable to conclude that the person in NY has high privacy loss because the published count is likely to help adversaries identify the person's location.
On the other hand, consider a person in Paris who is unlikely to be in NY. In such a case, it is implausible that the count from the red circle gives adversaries much information to identify the person's location---the person in Paris should have low privacy loss.

The disparity between the DP loss and intuitive privacy loss comes from the assumption of strong adversaries in DP who have arbitrary auxiliary knowledge.
More specifically, adversaries considered in DP contain the following adversary for the example above: an adversary knowing (1) all the locations of people except for the target person (illustrated with a star in Figure~\ref{fig:example1}), and (2) the target's location is either loc 1 or loc 2, where there are stars.
In such a case, publishing the count at the red circle will tell the target's location, whichever location the target is.
Due to this extreme assumption on the adversaries, the privacy loss is the same for people in NY, Paris, and anywhere in the world \emph{regardless of the activated sensor locations}.

An ad-hoc way to work around the pitfall of DP is to partition people into different datasets, e.g., based on the cities they live in. Note that the partition must be done such that each person belongs to only one dataset. However, this ad-hoc way is problematic especially when the sensors are incapable of knowing which dataset a person belongs to; for example, physical sensors on public roads counting the number of cars or cameras detecting and counting the number of people can hardly do so. Furthermore, partitioning people can result in publishing incorrect counts, particularly for locations where a lot of tourists visit. 
Instead, our work uses the relaxed adversary to better assess individual privacy loss without having such an ad-hoc workaround.

\subsubsection{Measuring Privacy Loss with Data Reconstruction Game~\cite{guo_analyzing_2022}}
Another way of measuring privacy loss in our setting is to directly evaluate the privacy loss of reconstruction attacks against a more relaxed adversary.
\cite{guo_analyzing_2022} propose such a framework to do so for discrete data.

In particular, the adversary in the proposed framework aims to identify their target's data $X$, which can take on $M$ discrete values $x_1,\ldots,x_M$. 
Similar to the DP adversaries, they have the data of all people except for the target's, $D^-$, as well as the (possibly randomized) mechanism output, $\mathcal{M}(D^-\cup {X})$.
However, instead of arbitrary auxiliary information that the DP adversaries have, the only additional auxiliary information they have is the \emph{prior distribution} of $X$, which is assumed to be a categorical distribution defined by the probability vector $\mathbf{p}$, i.e., $\Pr[X=x_m] = p_m$ for each $m \in \{1,\ldots,M\}$.

Then, the privacy loss against such an adversary is measured by the \emph{advantage} of the adversary. More concretely, let $\hat{X}$ be the guess of $X$ by the adversary and $p^* = \max_m p_m$, be the success rate of the Bayes optimal strategy, only with the prior information. Then, the advantage is defined as the normalized difference between $\Pr[\hat{X}=X]$ and $p^*$:
\begin{align*}
    \mathrm{Adv} = \frac{\Pr[\hat{X}=X]-p^*}{1-p^*},
\end{align*}
where the probability is taken over the randomness of \emph{both} $X$ and the mechanism $\mathcal{M}$. The advantage measures how much the adversary guesses more correctly by observing the output.
Furthermore, \cite{guo_analyzing_2022} provide the upper bound on the advantage by utilizing Fano's inequality.
More specifically, let $p_e = \Pr[\hat{X}\neq X]$ be the error probability for the adversary. Then, by Fano's inequality, it holds, regardless of the adversary's algorithm, that
\begin{align} \label{eq:fano-chuan}
    H(X) - I(X;\mathcal{M}(D^-\cup {X})) \leq H(p_e)+p_e\log(M-1).
\end{align}
Here, recall $X$ is the target's data generated from the known categorical distribution defined with the probability vector $p$. Thus, the entropy of $X$ is obtained easily by $H(X) = -\sum_m p_m\log p_m$.
Furthermore, $I(X;\mathcal{M}(D^-\cup {X}))$---the mutual information between $X$ and the mechanism output---is obtained through the analysis of the mechanism $\mathcal{M}$
~\footnote{Depending on $\mathcal{M}$, we obtain the upper bound of $I(X;\mathcal{M}(D^-\cup {X}))$. In such a case, we replace $I(X;Y)$ with its upper bound in Eq~\eqref{eq:fano-chuan}.}.
\cite{guo_analyzing_2022} further discuss the relationship between $I(X;\mathcal{M}(D^-\cup {X}))$ and DP mechanisms in their work.
Finally, using Equation~\eqref{eq:fano-chuan} as a constraint on $p_e$ and numerically finding its minimum (see Algorithm 1 in~\cite{guo_analyzing_2022} for the concrete algorithm), i.e., 
\begin{align*}
  p_e^* = \min\{p_e\in [0,1] : H(X)-I(X;\mathcal{M}(D^-\cup {X}))
  \leq H(p_e) + p_e\log(M-1)\},
\end{align*}
yield the lower bound on $p_e$, i.e., $p_e \geq p_e^*$ or equivalently the upper bound of $\Pr[\hat{X}=X]$, and thus, the advantage: 
\begin{align*}
    \mathrm{Adv} \leq \frac{1-p_e^* - p^*}{1-p^*}.
\end{align*}
Note that $p_e^*$ gets larger as the left hand side in eq~\eqref{eq:fano-chuan} becomes larger.

The framework fits into our setting by letting $X=X^1$, assuming the first person is the target of the adversary, $D^- = \{X^2=x^2,\ldots,X^N=x^N\}$, and $\mathcal{M}(D^-\cup {X})=(\mathrm{count}(D_1,c_1),\ldots,\mathrm{count}(D_T,c_T))$ be the publish counts across time.
Note that each person's data can take on $M^T$ discrete values.

Compared with DP, we see that 
the framework can take into account how a sensor in one location induces a different privacy loss to different individuals all over the world. 
Returning to Example 1, where $T=1$ for simplicity, we see that, for a person in Paris whose prior probability of being at any location in NY is very small, the count from the red circle in NY only has little information on $X$.
More specifically, assuming no noise is added to the count, we have $I(X;\mathrm{count}(D_1,c_1)) = H(\mathrm{count}(D_1,c_1)) = H(\Pr[X=\mathrm{loc 2}])$ which is small for the person in Paris. 
On the other hand, a person in NY who has a larger probability mass on the sensor location will have higher mutual information.
Therefore, when those two people have almost the same entropy of the prior, $\mathrm{Adv}$ is smaller, or privacy loss is smaller, for the person in Paris than the person in NY.
Moreover, the framework gives the semantically meaningful adversarial notion of privacy loss, which DP does not automatically offer.
Thus, we expect this framework to be quite appealing for practical applications.

However, there still needs much work to understand the privacy loss for our spatio-temporal counting problem under this framework mainly because we use temporal data.
First, the framework uses too strict a success definition---the adversary succeeds if they recover the target individual's locations at all time steps, i.e., $\hat{X}=X \iff \forall t.\quad \hat{X}_t = X_t$. It does not provide guarantees for more flexible and realistic success metrics, e.g., the adversary succeeding when they identify locations at only a few time steps.
Second, the framework assumes the person's data can take on $M^T$ values and is drawn from a general non-temporal categorical distribution.
Without further specifying the distribution, it becomes intractable to obtain the upper bound on the privacy loss as $T$ increases, e.g., computing $H(X)$ takes time scaling exponentially with $T$.
Fortunately, it is often the case for temporal data that distributions across time do not change rapidly.
Thus, we can assume a temporal relationship that links the distributions across time. By doing so, the privacy loss computation becomes tractable while still capturing the realistic privacy loss.
Furthermore, it enables us to explicitly observe how different temporal relationships offer different privacy loss as follows.
\paragraph{Example 2.}
\begin{figure}[t]
    \centering
    \includegraphics[width=\linewidth]{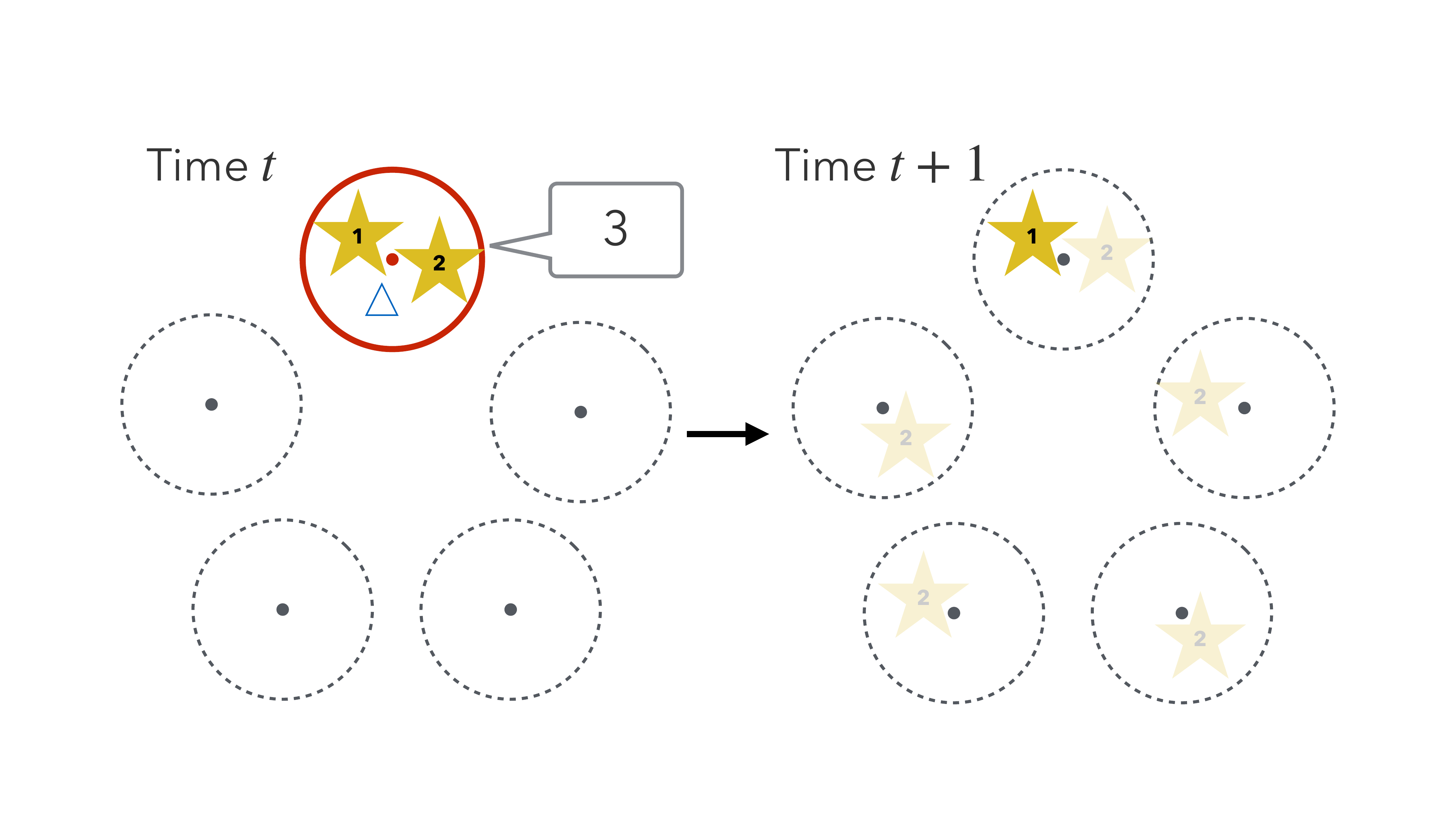}
    \caption{Our spatio-temporal counting problem at time $t$ and $t+1$. Each circle represents a location: a solid red circle indicates the count is published from the location, and dotted black circles indicate it is not. A blue triangle and yellow stars represent people. Since there are one triangle and two stars within the red circle at time $t$, the published count is 3. Person 1, who is labeled with "1", is likely to stay in the same location at the next time step. Person 2, who is labeled with "2", is likely to move to other locations at the next time step.}
    \label{fig:example2}
\end{figure}
Consider the spatio-temporal counting in Figure~\ref{fig:example2}. At time $t$, the number of people within the red solid circle is published, and at time $t+1$, the count at some location other than the five dotted circles is published. Person 1 and person 2, both illustrated with a star each labeled with "1" and "2" are within the red circle at time $t$. In the next time step, person 1 stays in the same location with a probability close to 1. By knowing person 1 is at the red circle at time $t$, the adversary can guess the person's location no matter which sensor gets activated at time $t+1$. 
On the other hand, person 2 can be any of the 5 circles at the next time step; thus, it is hard for the adversary to identify the person's location even when they know the person is at the red circle at time $t$.
Comparing person 1 and person 2, we expect that person 1 should have higher privacy loss than person 2 due to the heavy correlation across time.
We note that DP fails to capture this difference since DP considers the most slow-moving person, i.e., a person who stays in the same location at all time steps, and the privacy loss of any individual is the same as that of the worst-case person. 

Finally, the work does not demonstrate the impact of heterogeneous priors that reflect the spatial difference among individuals since its primary purpose is not to measure privacy loss in spatio-temporal data. Thus, in this work, we examine how spatio-temporal counting incurs different privacy loss for different individuals.

\section{Measuring Privacy Loss for Distributed Spatio-temporal Counting Problem}
\subsection{Our Framework}
We make the framework by \cite{guo_analyzing_2022} more tailored to measure privacy loss for distributed spatio-temporal counting problems.
In particular, we first make mild assumptions on our setting, and then specify a realistic adversary and suitable success metric.
In the rest of this paper, we use the raw counts, $(\mathrm{count}(D_1,c_1),\ldots,\mathrm{count}(D_T,c_T))$, as the output counts for readability unless otherwise noted.

\paragraph{Assumptions.}
We make the following assumptions about people's locations and how sensors at each location behave.
First, we assume $X^i$ forms a stationary Markov chain in time, i.e., $X^i_1\to X^i_2 \to \cdots \to X^i_T$ for any $i$. We let $\pi_i$ denote the initial prior and $P_i$ denote the transition matrix of $i$-th person. It is common in the literature~\cite{shokri_quantifying_2011,rafiei_quantifying_2022} to model temporal correlations over user locations with the Markov Chain. Moreover, the assumption still allows us to capture the privacy loss incurred by the temporal correlation as in Example 2. 
The second assumption is that locations $X^i$ for different individuals (different $i$) are independent. This is necessary to guarantee that the privacy loss for one person does not increase due to correlations between the person's locations and the others.
The third assumption is that only one sensor gets activated and counts the number of individuals at each time step, and the activated sensors are not adaptively chosen. 
We make this assumption for the sake of simplicity for our analysis. This is also validated by the fact that time is continuous in the real world. Evaluating privacy loss under adaptive sensor selection is one of our future work.

\paragraph{Adversary.}
In our framework, we consider a relaxed, yet realistic, adversary whose auxiliary information is not arbitrary. 
The goal of the adversary is to identify the target person's locations, $X$, as much as possible.
Without loss of generality, the target person is the first person, i.e., $X=X^1$. We use $\hat{X}=(\hat{X}_1,\ldots,\hat{X}_T)$ to denote the adversary's guess on the target person's locations.
Then, in addition to the published counts $(\mathrm{count}(D_1,c_1),\ldots,\mathrm{count}(D_T,c_T))$, they have the following information to compute $\hat{X}$: locations of all people except for the target, $D^- = \{X^2=x^2,\ldots,X^N=x^N\}$, and the prior of the target's locations, $\pi=\pi_1$ and $P = P_1$. 
We note that if the raw counts are published, the adversary is able to know whether the target is at the activated sensor or not at each time---it suffices for them to subtract the count of people except the target from the published count. However, they still have some uncertainty about the target's location if the target is not at the activated sensor.

Our specification of the adversary enables us to properly measure the privacy loss for our spatio-temporal counting---it allows us to assess how much information would be given to the adversary by the published counts depending on the target's prior.
For a concrete example, let us return to Example 1 (see Figure~\ref{fig:example1}).
Even if the published count is randomized, the adversary can correctly guess the location of a person in NY since the person is likely to be at the activated sensor location and the count may reduce the adversary's uncertainty a lot.
On the other hand, seeing the published count, the adversary cannot reduce their uncertainty on the location of a person in Paris because it is unlikely that the person is at the activated sensor.
Because of this distinction between people in NY and Paris, the probabilities that the adversary identifies each person's location should significantly differ; thus, our framework successfully measures appropriate individual privacy loss.

\paragraph{Definition of privacy loss.}
We measure the privacy loss by the success rate of the adversary's reconstruction attack or estimation. We set the definition of success to be more flexible and suitable for our setting.
In particular, we say the adversary succeeds when the number of time steps they guess incorrectly is no more than some threshold. To put it formally, let $d(\cdot,\cdot)$ be the hamming distance between locations. Then, the adversary succeeds if and only if $d(X,\hat{X}) \leq s$ for some $s (<T)$.
This is a strict generalization of the success definition used in the work by \cite{guo_analyzing_2022}---the adversary succeeds if and only if $\hat{X} = X$---because setting $s=0$ recovers their definition.
This generalized success metric provides a full picture of realistic privacy. For example, even leaking the exact location at one time step can be problematic. Furthermore, it is sometimes enough for the adversary to identify a part of the target's locations for inferring further on the target.
Our success definition makes it possible to evaluate the proper privacy loss depending on the specific situation.
Finally, we measure the privacy loss for our setting by the probability of the success event:
\begin{align}\label{eq:loss-def}
    \mathrm{Privacy Loss} = \Pr[d(X,\hat{X}) \leq s].
\end{align}
Note that the probability is taken over the randomness of $X$ and the additive noise if any. We further note that the privacy loss is defined for each adversarial estimator $\hat{X}$.
Our privacy loss differs from the one of \cite{guo_analyzing_2022} in that ours does not involve normalization with $p^*$, which is the success rate of the Bayes optimal strategy only with the prior information. This is because our success definition is not as simple as theirs; thus, it is intractable to obtain the exact value of $p^*$. 
However, under the same prior, our definition preserves the relative order of the privacy loss for different estimators.
Furthermore, an advantage of our privacy loss definition is that we are able to observe how different individual priors yield different privacy loss. For example, a prior having a high probability mass in one location yields a higher privacy loss than the uniform prior in general.

We now have the appropriate privacy loss measure for reconstruction attacks for the spatio-temporal counting problem. In the rest of this section, we present two ways of analyzing our privacy loss: (1) computing the privacy loss for specific adversarial estimators, and (2) theoretical upper bound on the privacy loss for \emph{any} adversarial estimators.

\subsection{Computation of Privacy Loss for Specific Adversarial Estimators}
\label{sec:specific-adversary}
Recall the goal of the adversary is to correctly guess the target's actual locations given the prior of the target's locations and the published counts.
Thus, for a single adversarial attack, they choose an \emph{estimator} of the target's locations and estimate the locations with it.
This only yields a binary result of whether the attack succeeds or not based on the success definition. However, we are interested in the probability of the success in Eq.~\eqref{eq:loss-def}.
Therefore, we provide a way to quantify our privacy loss for each specific adversarial estimator with the estimates.

\subsubsection{Computation of Privacy Loss From Estimates}
To obtain the approximation of our privacy loss for each adversarial estimator, we perform the Monte Carlo method.
More specifically, given the prior of the target locations, we repeat (1) sampling a trajectory $X=x=(x_1,\ldots,x_T)$ from the prior, (2) generating the raw counts or counts with additive noises, and (3) obtaining the estimate $\hat{X}$ using the adversarial estimator.
Then, we compute the success rate over multiple runs, yielding the unbiased estimate of the privacy loss for the adversarial estimator.

\subsubsection{Adversarial Estimators}
It remains to consider specific adversarial estimators to quantify our privacy loss.
Since we assume that the adversary has the prior distribution over the target locations, a maximum a posteriori (MAP) estimator is their natural choice. If the prior is correct, the MAP estimator achieves the best success rate, or the worst privacy loss, when the success is defined to identify the locations at all time steps ($s=0$).
In addition to the MAP estimator, we present two other baseline estimators: a maximum a priori, and a (lucky) constant estimator.

\paragraph{MAP estimator.}
Given the prior and output counts, the MAP estimator computes $\hat{X}^\mathrm{MAP}$ as follows:
\begin{align*}
    \hat{X}^\mathrm{MAP}
    &= \argmax_{x_1,\ldots, x_T} \Pr[X=x | \mathrm{count}(D_1,c_1),\ldots,\mathrm{count}(D_T,c_T)] \\
    &= \argmax_{x_1,\ldots, x_T} \prod_{t=1}^{T} \Pr[\mathrm{count}(D_t,c_t)| X_t = x_t]\cdot
    \prod_{t=1}^{T} \Pr[X_t=x_t | X_{t-1}=x_{t-1}].
\end{align*}

The estimation can be done efficiently in linear time and space in $T$.
\begin{proposition} \label{prop:MAP}
    There exists an algorithm that computes $\hat{X}^\mathrm{MAP}$ in $\mathcal{O}(TM^2)$ time and $\mathcal{O}(TM)$ space.
\end{proposition}

\paragraph{Maximum a priori estimator.}
The second estimator is the maximum a priori estimator, which guesses the most probable locations given only the prior---the estimator does not use the published counts.
This estimator serves as a baseline of how much the prior can leak the target's locations.
In particular, this estimator outputs:
\begin{align*}
    \hat{X}^\mathrm{prior} = \argmax_{x} \Pr[X=x].
\end{align*}
This is done in linear time and space in $T$.
\begin{proposition} \label{prop:prior}
    There exists an algorithm that computes $\hat{X}^\mathrm{prior}$ in $\mathcal{O}(TM^2)$ time and $\mathcal{O}(TM)$ space.
\end{proposition}

\paragraph{Lucky constant estimator.}
The last estimator we consider is the (lucky) constant estimator, which guesses one of the locations for all time steps without seeing either of the prior or published counts.
More formally, the estimator outputs $\hat{X}^\mathrm{constant}=(l,\ldots,l)$ for some $l\in [M]$ with probability $1$. 
The privacy loss for this estimator is expected to be low in general. However, when the target's locations correlate heavily and do not change over time, the privacy loss can be high enough; thus, it serves as a good baseline for such occasions.

\subsection{Upper Bound on Privacy Loss for Any Adversarial Estimator}
While the Monte Carlo method provides an unbiased estimate of our privacy loss for each adversarial estimator, it requires massive computation time for accurate estimates. Therefore, we present two information-theoretic upper bounds on the privacy loss for \emph{any} adversarial estimator. Although  the privacy loss upper bounds provide conservative estimates of our privacy loss, they depend on individual's prior distribution, and thus still provide more realistic and fine-grained measure of privacy compared to DP, which is often overly conservative for most individuals in our setting.

To obtain the upper bound of our privacy loss, we follow the idea from \cite{guo_analyzing_2022} of using Fano's inequality, which in our case relates the error rate, or equivalently the success rate, of guessing the individual's locations to how much information of the locations the published counts retain. 
However, since our success definition is more general than the one used in \cite{guo_analyzing_2022}, we cannot directly use Fano's inequality in eq.~\eqref{eq:fano-chuan} to obtain the privacy loss upper bound.
Instead, we use variants of Fano's inequality, and derive the bounds for the spatio-temporal setting.

\subsubsection{Upper bound on privacy loss from Fano-type inequality}
\label{sec:classical-ub}
Recall that under our success definition, the adversary succeeds if and only if the number of incorrect guesses is below some threshold.
Thus, we use the variant of Fano's inequality which is more general and relaxes the error, or success, definition to ours to obtain the privacy loss upper bound.
In particular, let $E = \textbf{1}[d(X, \hat{X}) > s]$ be the error event indicator variable. Then, we use the following variant of Fano's inequality~\cite{duchi_distance-based_2013}:
\begin{align}\label{eq:fano-ours}
     H(X) - I(X;\mathrm{count}(D_1,c_1),\ldots,\mathrm{count}(D_T,c_T))
     \leq H(E) + \Pr[E=1] \log \frac{M^T-N(s)}{N(s)} + \log N(s),
\end{align}
where $N(s) = \sum_{l=0}^s \genfrac(){0pt}{2}{T}{l} (M-1)^l$ is the cardinality of $X$ given the success event happens. 
We first compute the entropy and mutual information in eq.~\eqref{eq:fano-ours} and then find the maximum possible value of the success probability satisfying the inequality, which is exactly our privacy loss upper bound.

\paragraph{Computing $H(X)$.}
Since we assume that $X$ forms the Markov Chain, we have $H(X) = H(X_1) + \sum_{t=2}^{T} H(X_t|X_{t-1})$. This is computed easily with the prior, $\pi$ and $P$.

\paragraph{Computing mutual information.}
The remaining difficulty in eq.~\eqref{eq:fano-ours} is to compute the mutual information $I(X;\mathrm{count}(D_1,c_1),\ldots,\mathrm{count}(D_T,c_T))$. 
In fact, obtaining the exact value of it is intractable even when the raw counts are published. The computation time scales exponentially with $T$.
Therefore, instead of the exact computation, we use the upper bound on $I(X;\mathrm{count}(D_1,c_1),\ldots,\mathrm{count}(D_T,c_T))$ and write $\tilde{I}$ to denote it, i.e., $I(X;\mathrm{count}(D_1,c_1),\ldots,\mathrm{count}(D_T,c_T)) \leq \tilde{I}$.

When we publish the raw counts, we use $\tilde{I} = \sum_{t=1}^T H(\mathrm{count}(D_t,c_t)|\mathrm{count}(D_{t-1},c_{t-1}))$ for the upper bound on the mutual information.~\footnote{We could increase the number of conditional variables to obtain tighter bounds, but there's a tradeoff between tightness and computation.} 

When the Gaussian noise is added to the raw count at each time, i.e., the Gaussian mechanism~\cite{dwork_algorithmic_2014} for DP is used, we use another upper bound:
\begin{align}\label{eq:MI-Gaussian}
    &I(X;\mathrm{count}(D_1,c_1),\ldots,\mathrm{count}(D_T,c_T))\nonumber\\
    &\leq \sum_{t=1}^{T}
    -p_t\log\left(p_t + (1-p_t)\exp\left(\frac{-1}{2\sigma^2}\right)\right)
    - (1-p_t)\log\left((1-p_t) + p_t\exp\left(\frac{-1}{2\sigma^2}\right)\right),
\end{align}
where $\sigma^2$ is the variance of the Gaussian noise and $p_t = \Pr[X_t=c_t]$. 
This follows by the slight modification of Theorem 2 in~\cite{guo_analyzing_2022}.
Then, we set the right-most side of eq.~\eqref{eq:MI-Gaussian} as $\tilde{I}$.

\paragraph{Obtaining upper bound on privacy loss.}
Given $H(X)$ and $\tilde{I}$, we obtain the privacy loss upper bound through the same procedure done by~\cite{guo_analyzing_2022}.
In particular, since it holds for any adversarial estimator that
\begin{align*}
    H(X) - \tilde{I} \leq H(E) + \Pr[E=1] \log \frac{M^T-N(s)}{N(s)} + \log N(s),
\end{align*}
we use the inequality as a constraint on $p_e = \Pr[E=1]$ and numerically find its minimum, i.e.,
\begin{align*}
    p_e^* = \min\biggl\{p_e\in [0,1]:
    H(X) - \tilde{I}
    \leq H(p_e) + p_e \log \frac{M^T-N(s)}{N(s)} + \log N(s)\biggr\}.
\end{align*}
The privacy loss bound is then: $\Pr[d(X,\hat{X}) \leq s] \leq 1 - p_e^*$. 

\subsubsection{Tighter upper bound on privacy loss for non-uniform priors}
\label{sec:tighter-ub}
The upper bound obtained via eq.~\eqref{eq:fano-ours} is known to be tight when the prior distribution of $X$ is uniform. 
However, in our spatio-temporal setting where $X=(X_1,\ldots,X_T)$ is a location trace, the distribution is rarely uniform, leading the bound to be less meaningful.
To this end, we use another variant of Fano's inequality, which looks at the distribution of $X$ more carefully, and 
provide a tighter bound on our privacy loss when the prior of $X$ is non-uniform.

Again, let $E = \textbf{1}[d(X, \hat{X}) > s]$ be the error event indicator variable.
By generalizing the variant of Fano's inequality from~\cite{han_generalizing_1994} to our success metrics, we have the following inequality:
\begin{align*}
    - I(X;\mathrm{count}(D_1,c_1),\ldots,\mathrm{count}(D_T,c_T))
    \leq H(E) + \Pr[E=0] \log\max_{\hat{x}} \Pr[X\in\{x:d(x,\hat{x}) \leq s\}].
\end{align*}

To follow the same procedure for obtaining the privacy loss upper bound in Section~\ref{sec:classical-ub}, we need to compute $\max_{\hat{x}} \Pr[X\in\{x:d(x,\hat{x}) \leq s\}]$ or its upper bound.
While the computation takes more time than the entropy computation required in Section~\ref{sec:classical-ub}, we show it can still be computed with polynomial time in $T$ respectively for (1) $s=0$, i.e., $E=\textbf{1}[X\neq \hat{X}]$ (exact), and (2) $s > 0$ (upper bound).

When $s=0$, i.e., the adversary succeeds only when they guess locations at all time steps correctly, it holds that $\max_{\hat{x}} \Pr[X\in\{x:d(x,\hat{x}) \leq s\}] = \max_{x} \Pr[X=x]$.
We obtain this in linear time in $T$ as in Section~\ref{sec:specific-adversary}.

When $s>0$, we first provide the upper bound on $\max_{\hat{x}} \Pr[X\in\{x:d(x,\hat{x}) \leq s\}]$:
\begin{align} \label{eq:ub-for-tight-fano}
    \max_{\hat{x}} \Pr[X\in\{x:d(x,\hat{x}) \leq s\}]
    \leq 
    \sum_{l=0}^s \genfrac(){0pt}{2}{T}{l} 
    \max_{1\leq t_1 < \cdots < t_{T-l}\leq T} \max_{\hat{x}_{t_1},\ldots,\hat{x}_{t_{T-l}}} \Pr[X_{t_1}=\hat{x}_{t_1},\ldots,X_{t_{T-l}}=\hat{x}_{t_{T-l}}].
\end{align}

Then, it remains to compute each term in the summation on the right-hand side of eq.~\eqref{eq:ub-for-tight-fano}.
We show that it is possible to obtain it in polynomial time and space in $T$.

\begin{theorem} \label{thm:max_prob}
    There exists an algorithm that computes the right-hand side of eq.~\eqref{eq:ub-for-tight-fano} in $\mathcal{O}(T^3M^2 + T\log TM^3)$ time and $\mathcal{O}(TM^2)$ space.
\end{theorem}

Finally, we obtain the privacy loss upper bound with the same procedure in Section~\ref{sec:classical-ub}.

\section{Experiment}
We empirically investigate how our framework can successfully measure meaningful individual privacy loss. In particular, we ask the following questions:
\begin{enumerate}
    \item Is privacy loss higher for people who are more frequently observed by the sensors (Example 1)?
    \item Is privacy loss higher for people who move more slowly (Example 2)?
    \item How does privacy loss vary with parameters such as the number of time steps ($T$) and the number of locations ($M$)?
    \item What is the privacy loss of publishing raw counts compared to noisy counts with different levels of noise?
    \item How loose are our privacy loss upper bounds?
\end{enumerate}
We answer the first two questions with real datasets in Section~\ref{sec:q1}. We address the third question in Section~\ref{sec:q2} and the fourth question in Section~\ref{sec:q3} with simulated and real datasets. The fifth question is addressed both in Sections~\ref{sec:q2} and~\ref{sec:q3}.

\subsection{Methodology}
\paragraph{Data.}
For simulated data, we consider 1-dimensional discrete locations. 
In particular, we have $M$ locations, where each takes a different value in $\{1,\ldots,M\}$. 
We define the transition probability to be 
$\Pr[X_{t+1}=x_{t+1} | X_{t}=x_{t}] \propto \exp ( -\frac{|x_{t+1}-x_{t}|}{\tau M})$ with a parameter $\tau$ (the larger $\tau$ is, the more independent the locations are across time), and set the stationary distribution of the transition matrix as the initial prior.
In the experiments, we sweep parameters to study their effects. When not swept, we fix $M=100$, $T=10$, $\tau=0.1$, and $s=5$, i.e., adversary's success is defined to be $d(X,\hat{X})\leq s=5$.

For real data, we use Foursquare dataset~\cite{yang_nationtelescope_2015, yang_participatory_2016}, which contains check-in information associated with user ID from the location data platform Foursquare, and Gowalla dataset~\cite{cho_friendship_2011}, which also contains check-in information from a location-based social networking website called Gowalla.
We treat a time window of $20$ days as one time step, which makes the number of time steps to be $27$ for Foursquare and $32$ for Gowalla. Furthermore, we choose the most visited $100$ locations, and select a subset of users who have check-ins to the chosen locations at least $10$ time steps. This yields the number of users to be $N=9760$ for Foursquare and $N=318$ for Gowalla. For time steps that a user does not have any check-ins, we set the "somewhere else" location to be the checked-in location of the user.
Thus, the number of locations is $M=101$.
We then estimate the transition matrix for each user by the check-ins at the time steps except for the last $5$ time steps. Moreover, we set the stationary distribution as the initial prior.
Then, we use the remaining $T=5$ to be the actual trajectory that the adversary aims to identify. When not swept, we set $s=1$, i.e., the adversary's success is defined to be $d(X,\hat{X})\leq s=1$.

Recall we let $D_t = \{X^i_t: i=1,\ldots,N\}$ be the people's locations and $c_t$ be the sensor location which turned on at time $t$.
We choose the sensor locations uniformly at random at each time step unless otherwise noted.
Then, we publish either the raw counts, $\mathrm{count}(D_t,c_t)$ for all $t$, or the randomized counts with the Gaussian mechanism with standard deviation parameter $\sigma$, $\mathrm{count}(D_t,c_t) + \xi_t$ for all $t$, where $\xi_t \sim \mathcal{N}(0,\sigma^2)$.
We fix $\sigma=1$ except when sweeping it.

\paragraph{Experiment Setup.}
We compare our privacy loss upper bounds in Sections~\ref{sec:classical-ub} (\textbf{Loose UB}) and~\ref{sec:tighter-ub} (\textbf{Tight UB})
with the privacy loss for three adversarial estimators: the \textbf{MAP estimator}, the maximum a priori estimator (\textbf{Prior estimator}), and the lucky constant estimator (\textbf{Constant estimator}).
For the lucky constant estimator, we compute the privacy loss $M$ times~\footnote{We exclude the "somewhere else" location as a target for real datasets.}. For the $i$-th time computation, we compute the privacy loss for the estimator which always outputs $i$-th location at all time steps.
Then, we report the maximum over the $M$ privacy loss as the final privacy loss for the lucky constant estimator.

When we compute the privacy loss for the three adversarial estimators with simulated data by the Monte Carlo (MC) method, we generate $1000$ independent trajectories from priors, and average the success rates over these runs. 
For real datasets, instead of generating trajectories from estimated priors, we use real trajectories---the adversary guesses the locations of each individual with $T=5$, which are not used for the transition matrix estimation. We then report the attack success rates over all individuals which is an approximation of the mean individual privacy loss.
To compute the privacy loss upper bounds in Sections~\ref{sec:classical-ub} and~\ref{sec:tighter-ub}, we use the defined transition matrix for simulated data and use the estimated transition matrix for real datasets. For real datasets, we report the mean loss upper bounds over individuals.

\subsection{Results and Discussion}
\subsubsection{Intuitiveness of Privacy Loss} \label{sec:q1}
\begin{figure}[t]
\begin{subfigure}[b]{0.48\textwidth}
         \centering
         \includegraphics[width=\textwidth]{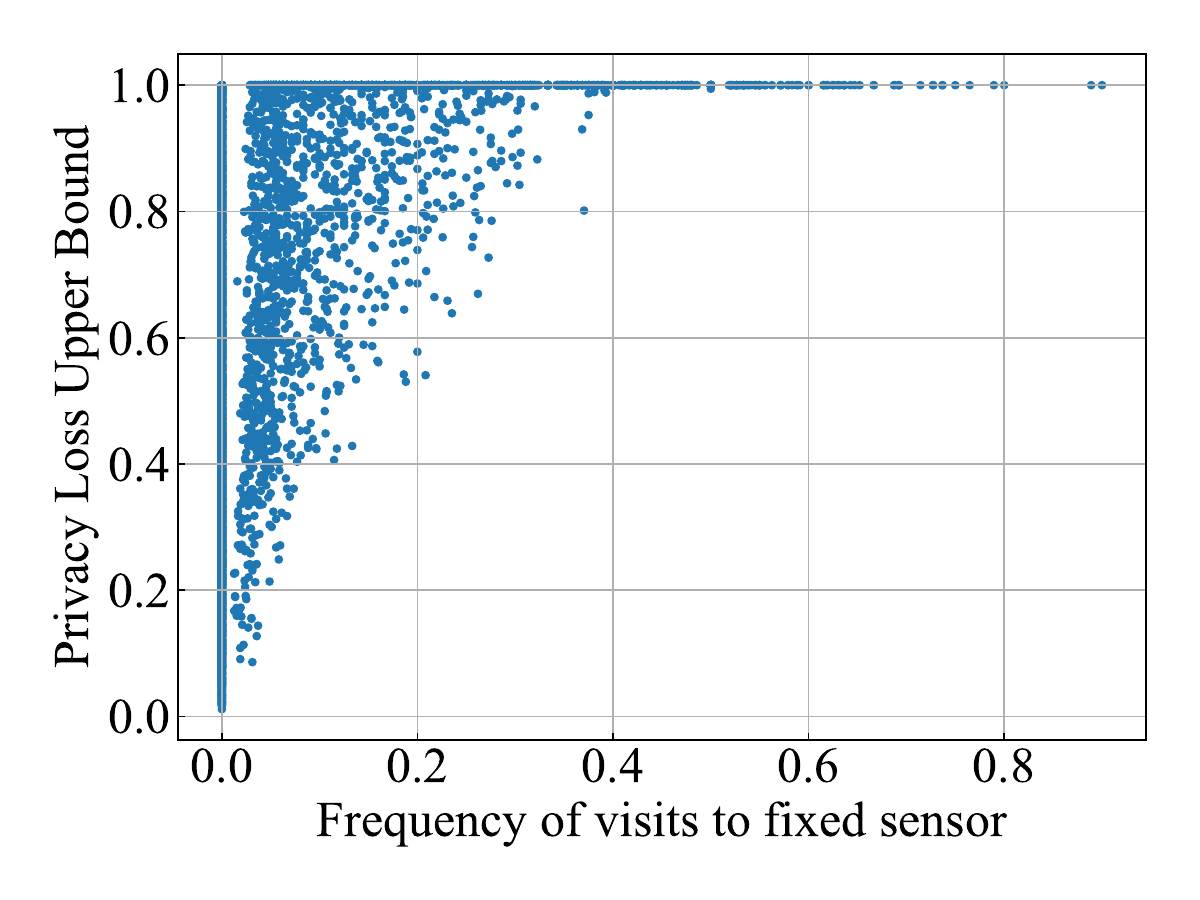}
         \caption{Foursquare}
     \end{subfigure}
     \hfill
     \begin{subfigure}[b]{0.48\textwidth}
         \centering
         \includegraphics[width=\textwidth]{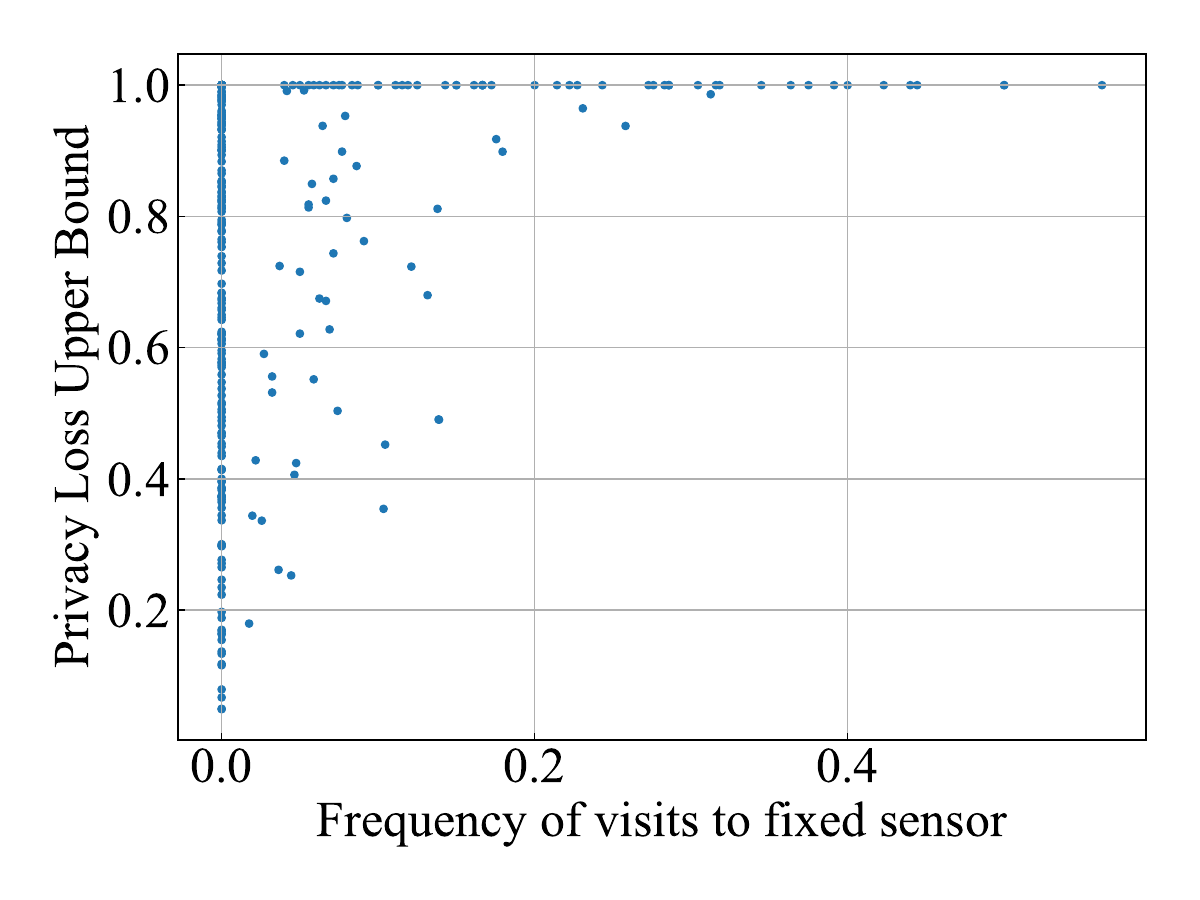}
         \caption{Gowalla}
     \end{subfigure}
    \caption{Relationship between the privacy loss upper bound in Section~\ref{sec:tighter-ub} and the frequency of visits to fixed activated sensor location of each individual on Foursquare and Gowalla datasets. Note that for this plot, we choose and fix a single sensor to be activated. We observe those who have visited the fixed sensor location frequently have high privacy loss. The privacy loss for those who haven't visited the fixed sensor much is scattered between 0 and 1 since the privacy loss depends on other factors as well.}
  \label{fig:exp-ex1}
\end{figure}
\begin{figure}[t]
\begin{subfigure}[b]{0.48\textwidth}
         \centering
         \includegraphics[width=\textwidth]{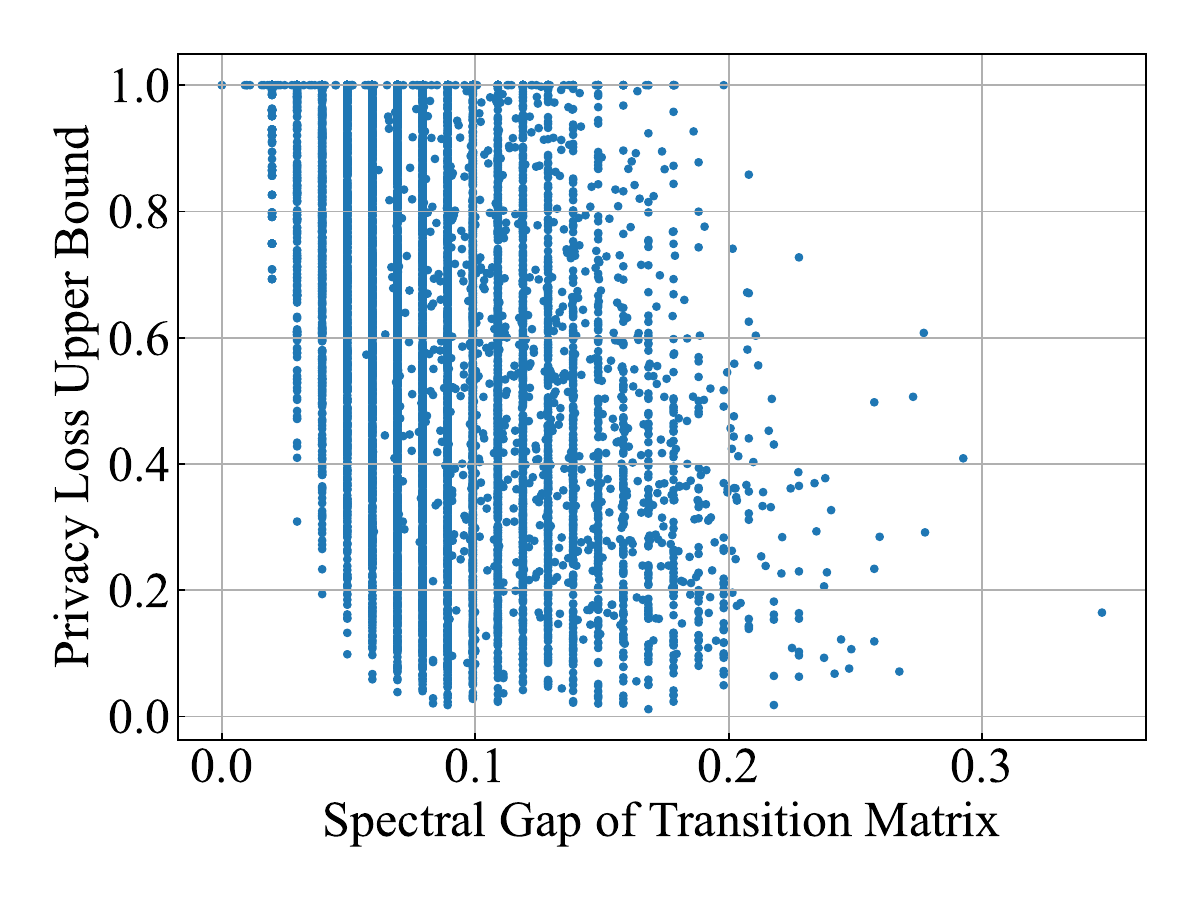}
         \caption{Foursquare}
     \end{subfigure}
     \hfill
     \begin{subfigure}[b]{0.48\textwidth}
         \centering
         \includegraphics[width=\textwidth]{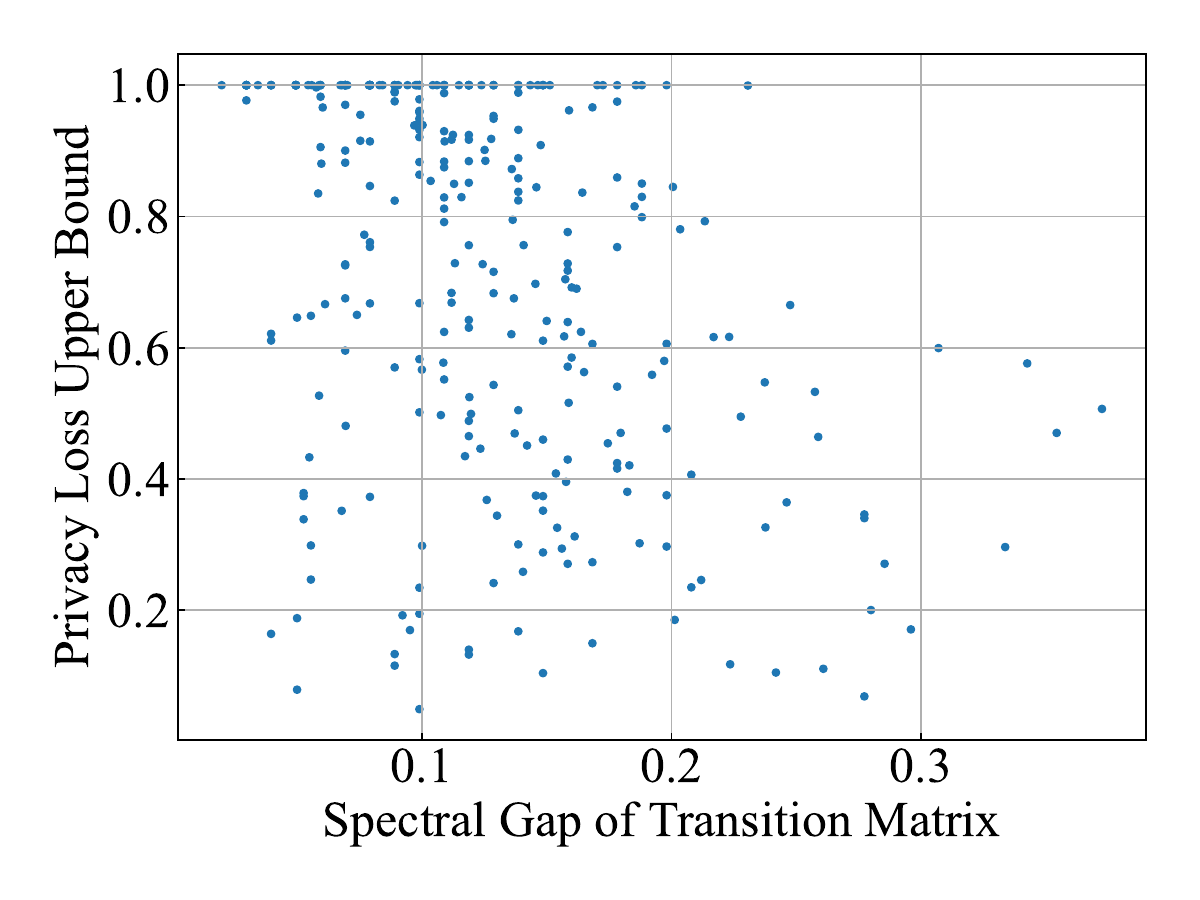}
         \caption{Gowalla}
     \end{subfigure}
    \caption{Relationship between the privacy loss upper bound in Section~\ref{sec:tighter-ub} and the spectral gap of each individual's transition matrix on Foursquare and Gowalla datasets. Note that a higher spectral gap of transition matrix means that the Markov Chain induced by the person's trajectory mixes rapidly---namely, the person is fast-moving. We observe the negative correlation between the privacy loss and the spectral gap. Since the privacy loss depends on other factors as well, there are people whose transition matrices have small spectral gaps, while their privacy loss is small.}
  \label{fig:exp-ex2}
\end{figure}
We first examine if our privacy loss reflects the intuitions described in Examples 1 and 2.

Recall our intuition described in Example 1 is that people likely to be observed more by sensors have higher privacy loss. 
To this end, we observe the individual privacy loss when the activated sensor location is fixed to be the same at all time steps. We choose the activated sensor location to be the most commonly visited location.
Figure~\ref{fig:exp-ex1} shows how each individual's privacy loss differs by their likeliness of being at the activated sensor location on the Foursquare and Gowalla datasets. For both datasets, we observe that individuals who have visited the fixed location frequently have high privacy loss. The results partially validate our intuition. On the other hand, those who have never visited the activated sensor location can have higher privacy loss. This is because the privacy loss depends on other factors such as individual priors. The adversary might be able to guess the target's locations only with their prior even when the target is not observed by sensors.

Figure~\ref{fig:exp-ex2} shows the relationship between the privacy loss and the spectral gap of the transition matrix for each individual on the Foursquare and Gowalla datasets.
In general, a larger spectral gap means that the Markov Chain induced by the person's trajectory mixes rapidly, or put simply, the person is fast-moving. Thus, we expect that the larger spectral gap yields less privacy loss.
Indeed, this is what we generally observe, and the results validate our intuition in Example 2 that fast-moving people, who have more uncertainty of their location given the previous location, should have smaller privacy loss. However, since the loss can be small or large due to other reasons, some individuals do not follow the overall trend.
We further argue that the results exhibit the advantage of our framework of measuring the privacy loss over the one of \cite{guo_analyzing_2022}. Our framework appropriately models the individual's locations with the Markov Chain; thus, we can explicitly investigate the relationship between the spectral gap, or how fast-moving the person is, and the privacy loss while keeping the privacy loss computation tractable.

Although the results do not validate our intuitions perfectly, we see a clear distinction between our privacy loss and the DP loss. Under DP, every person has the same privacy loss regardless of how likely they are observed by the sensors and how fast-moving they are. In particular, everyone has the loss of the worst-case individual such as the one who is always observed by the sensors or the one who stays in the same location at all times. In contrast, we see that our framework gives different privacy loss to each person based on multiple intuitive factors.

Furthermore, we note that we quantify our privacy loss of individuals when we publish the true counts over time---we do not add any noise to the counts. Therefore, the privacy loss in DP becomes infinite for all individuals in this setting. Even if we add noise to counts, the privacy loss is uniform across individuals. This fact supports that our privacy loss is a more fine-grained measure of individual privacy loss than DP.

\subsubsection{Effect of Parameters on Privacy Loss} \label{sec:q2}
\begin{figure}[t]
\begin{subfigure}[b]{0.48\textwidth}
         \centering
         \includegraphics[width=\textwidth]{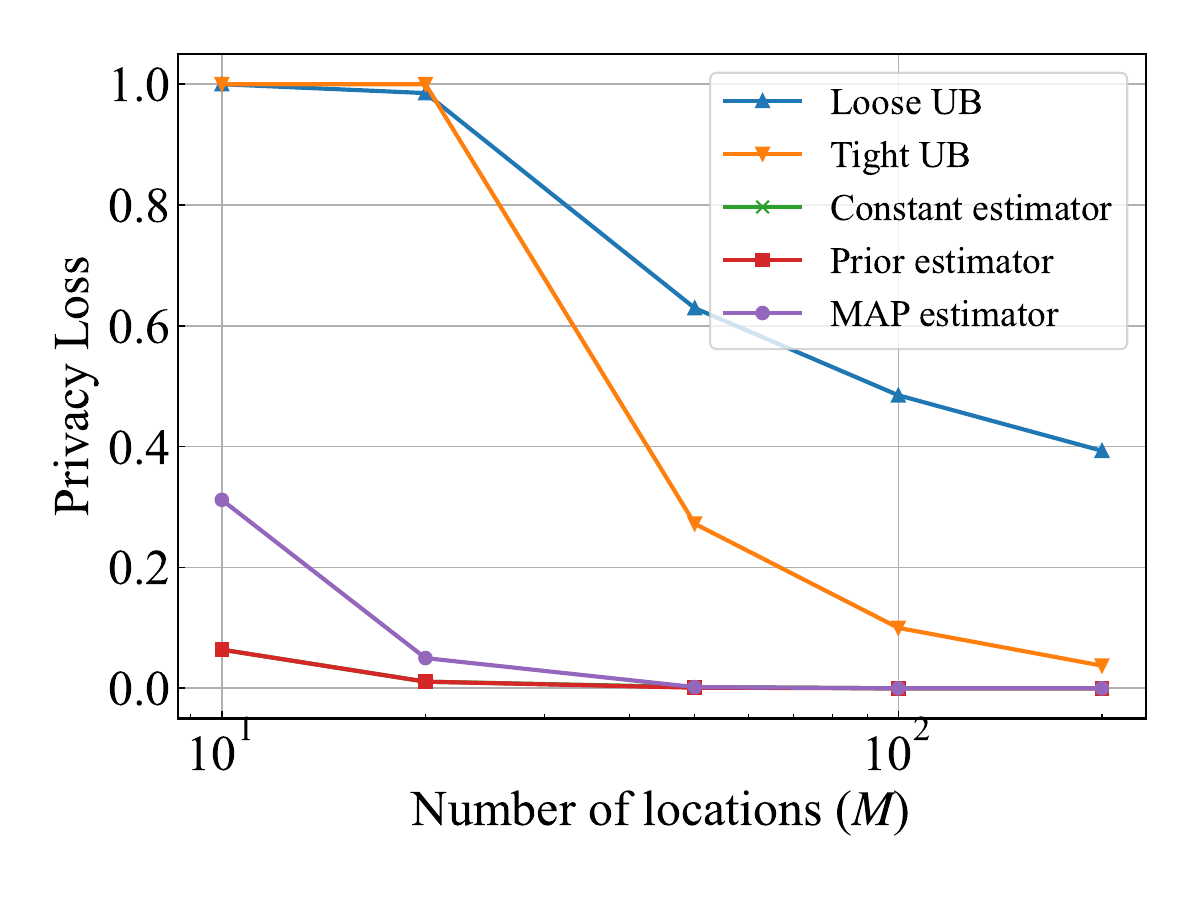}
     \end{subfigure}
     \hfill
     \begin{subfigure}[b]{0.48\textwidth}
         \centering
         \includegraphics[width=\textwidth]{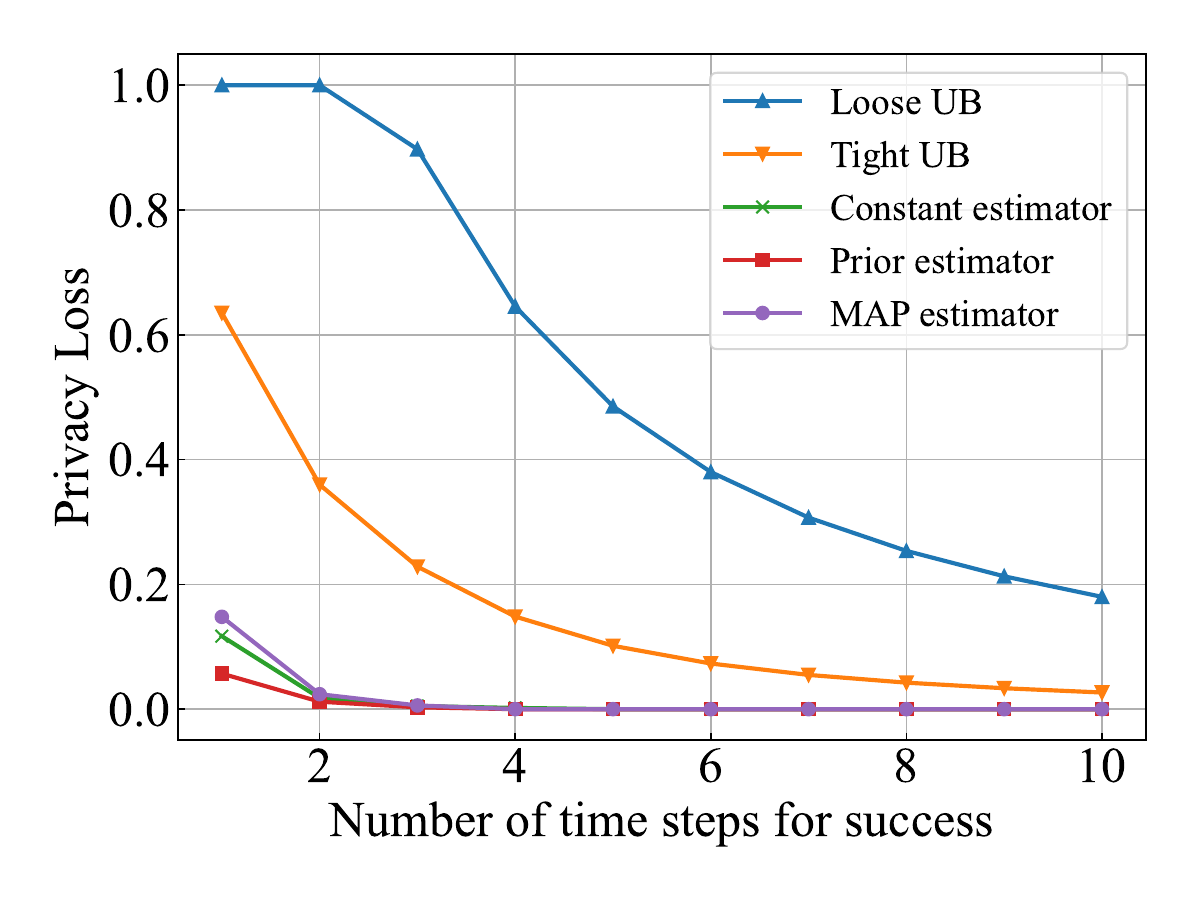}
     \end{subfigure}
     \hfill
     \begin{subfigure}[b]{0.48\textwidth}
         \centering
         \includegraphics[width=\textwidth]{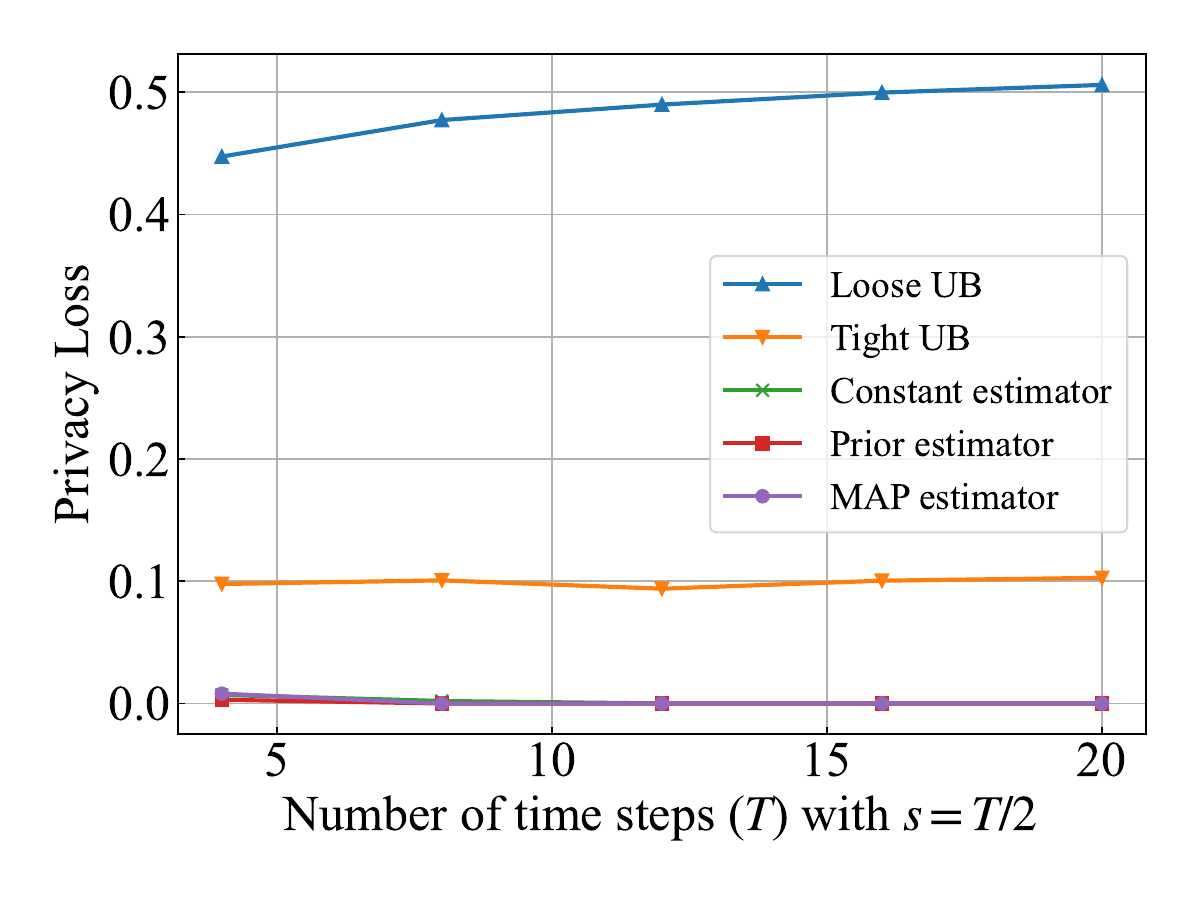}
     \end{subfigure}
     \hfill
     \begin{subfigure}[b]{0.48\textwidth}
         \centering
         \includegraphics[width=\textwidth]{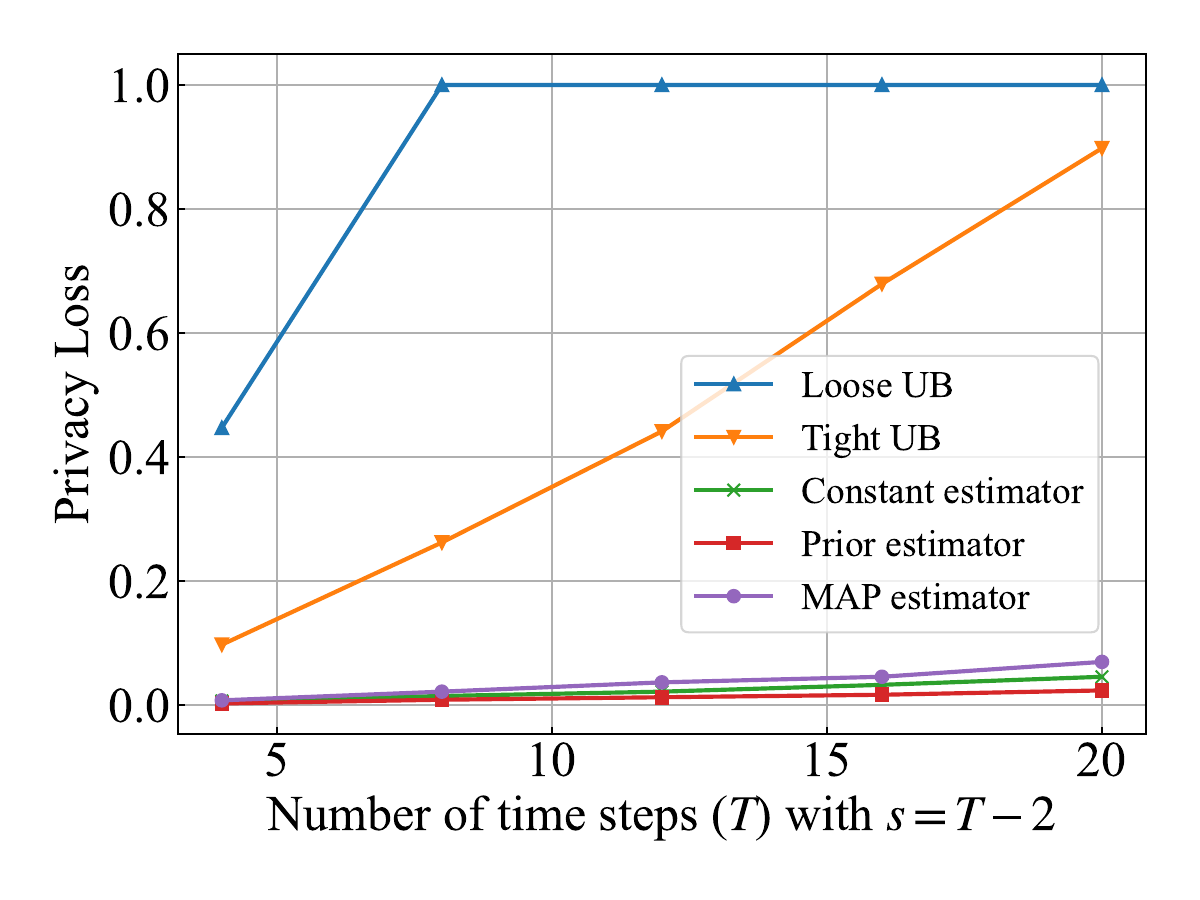}
     \end{subfigure}
    \caption{Privacy loss for the adversarial estimators and upper bounds on the simulated data. We sweep parameters: the number of locations $M$ (upper left), the number of time steps the adversary needs to guess correctly for success (upper right), the number of time steps $T$ with $s=T/2$ (lower left), and $T$ with $s=T-2$ (lower right). The privacy loss generally decreases as $M$ and the number of time steps for the adversary's success grow. As $T$ grows, it stays almost the constant when $s=T/2$ and increases when $s=T-2$.}
  \label{fig:exp-sim-params}
\end{figure}
\begin{figure}[t]
\begin{subfigure}[b]{0.48\textwidth}
         \centering
         \includegraphics[width=\textwidth]{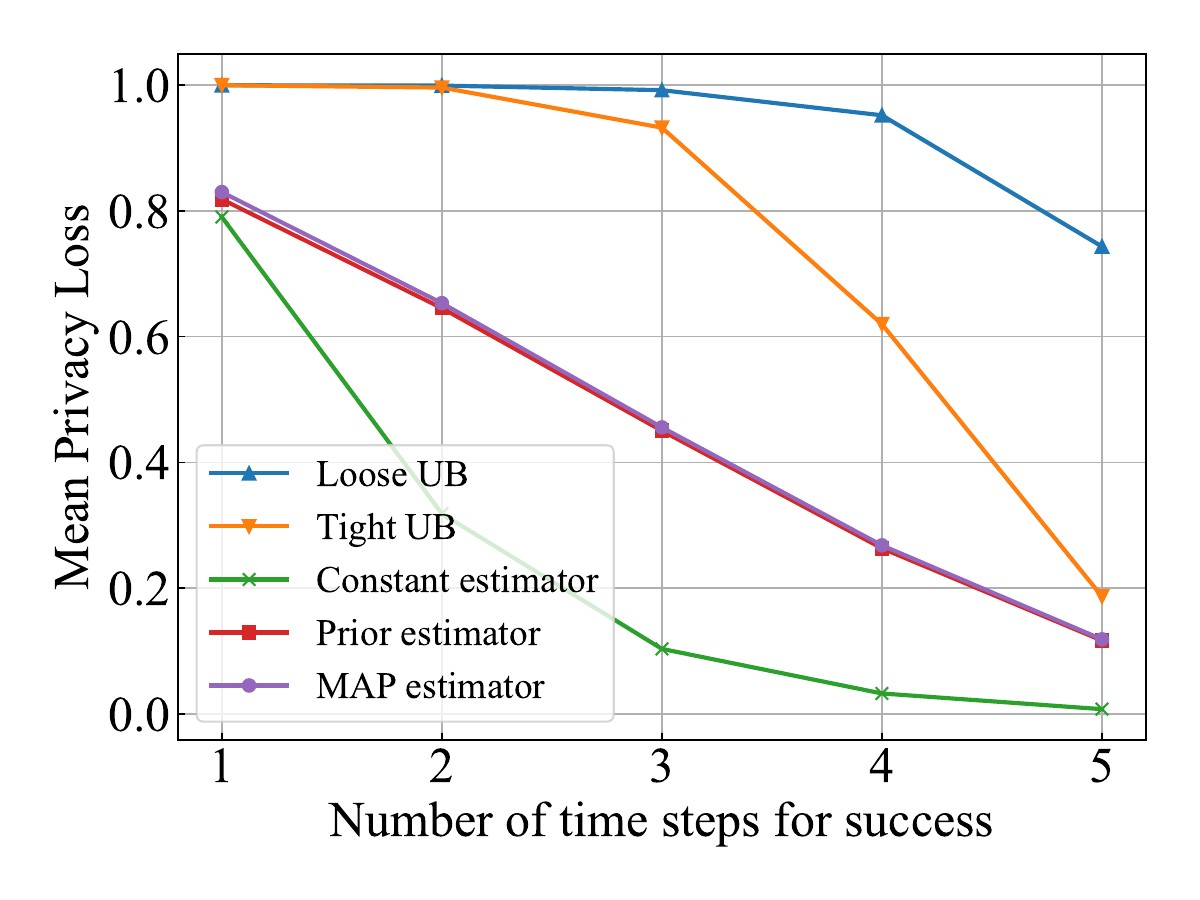}
         \caption{Foursquare}
     \end{subfigure}
     \hfill
     \begin{subfigure}[b]{0.48\textwidth}
         \centering
         \includegraphics[width=\textwidth]{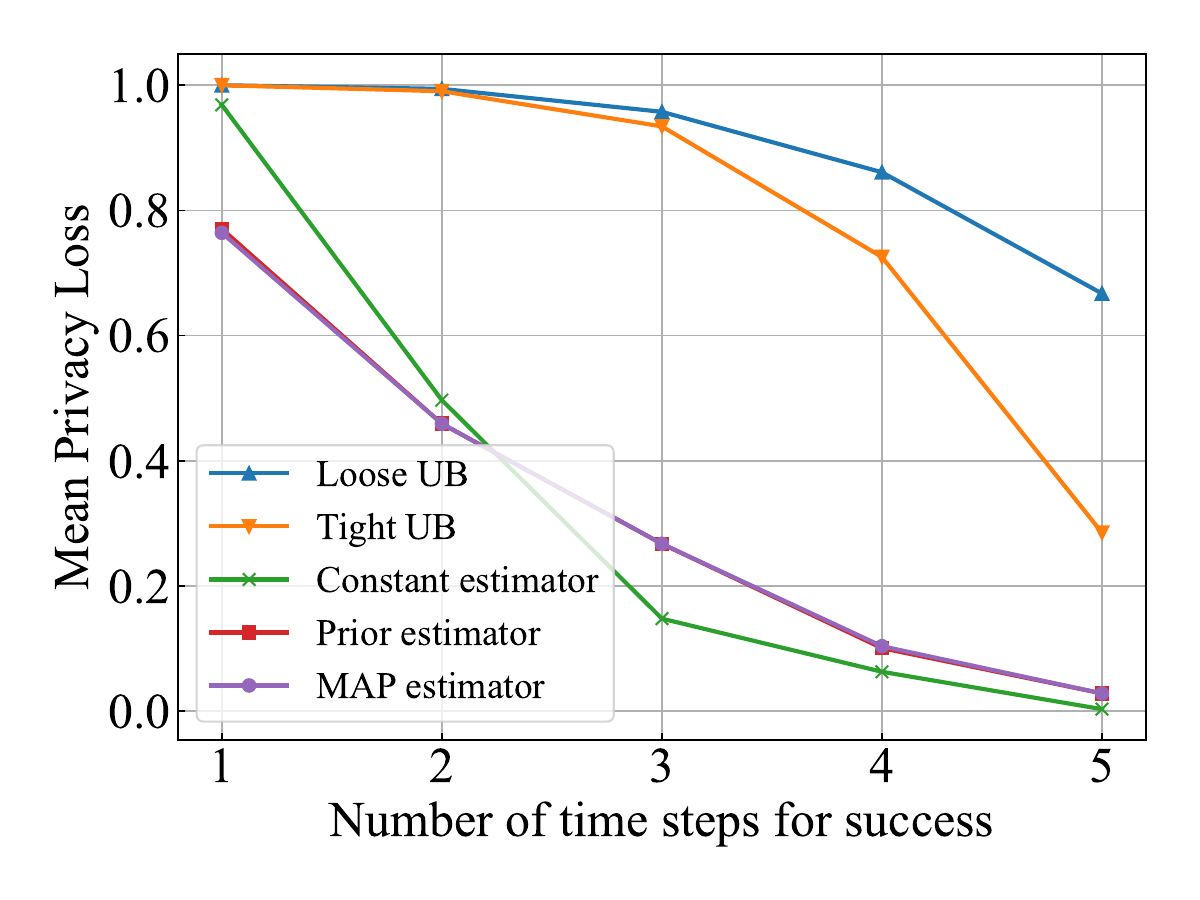}
         \caption{Gowalla}
     \end{subfigure}
    \caption{Mean privacy loss for the adversarial estimators and upper bounds over individuals with varying number of time steps for the adversary's success on the Foursquare and Gowalla datasets. The mean privacy loss decreases as the number of time steps for the adversary's success increases.}
  \label{fig:exp-real-params}
\end{figure}
Figure~\ref{fig:exp-sim-params} shows the comparison of our privacy loss for the adversarial estimators and upper bounds when the number of locations ($M$), the number of time steps the adversary can wrongly guess for a successful attack ($s$), and the number of time steps ($T$) are swept respectively on the simulated data.

We find that the privacy loss decreases as the number of locations $M$ gets larger. This is because, under larger $M$, individual's locations have higher uncertainty in general---the adversary has more choices to guess. We also observe the upper bounds get tight when $M$ is large, which happens in more realistic settings. We also see that the privacy loss for MAP is high $\approx 30\%$ for small $M$, indicating the hardness of preserving privacy in such a case. While we observe the change in privacy loss under different numbers of locations, this is not the case for DP---DP requires us to add the same level of noise and gives the same privacy loss for any number of locations.

As the number of time steps for success gets larger ($s$ is smaller), the privacy loss gets smaller as expected. The gap between the two bounds is large, and the tight UB provides a meaningful bound even when the loose one does not. Furthermore, tight UB is especially tight and almost matches the privacy loss for adversarial estimators when $s$ is small, i.e., the adversary needs to guess more time steps correctly.
The privacy loss for each estimator also decreases with smaller $s$. In addition, when the number of time steps for success is $1$, the lucky constant estimator yields a privacy loss. This indicates that for such a success metric, even the simple rule-based estimator can have a non-negligible chance to estimate the locations correctly.
The advantage of our privacy loss over DP and the one of \cite{guo_analyzing_2022} is that we can observe privacy loss against reconstruction attacks under more flexible success definitions. The definition of \cite{guo_analyzing_2022} only allows us to see the privacy loss when $s=0$, i.e., the right-most plots in the figure.

To investigate the effect of the number of time steps, we consider two cases: $s$ scales with $T$, here $s=T/2$, and the number of time steps for success remains fixed, here $s=T-2$. For the first case, the privacy loss remains almost constant across $T$'s because the number of time steps the adversary needs to guess correctly scales by the same rate as $T$ grows. On the other hand, for the second case, the privacy loss increases as $T$ grows. This is due to correlation across time, i.e., the adversary gets more information to guess locations at a fixed number of time steps correctly by having more time steps. The tightness of the upper bounds is the best when $T$ is the smallest, which suggests the difficulty of preventing attacks when the adversary only aims to identify the locations at a few time steps out of large $T$.
In DP, increasing the number of time steps gives larger $\epsilon$ if we fix the additive noise standard deviation to be constant. Our privacy loss has more flexibility in that we can measure how the privacy loss changes by the number of time steps under different adversary's goals by appropriately setting the parameter $s$.

Figure~\ref{fig:exp-real-params} demonstrates the comparison of the privacy loss for adversarial estimators and upper bounds on the real datasets. Note that the number of time steps for success is the only parameter we can sweep. We further note that we report the mean privacy loss for real datasets.
We observe almost the same trend as the one from the simulated data that the privacy loss diminishes as the number of time steps for success gets larger. However, we see the adversarial attacks succeed more for the real datasets. This suggests that the adversary can identify people's locations more easily in the real-world setting, even with a very simple attack such as the lucky constant estimator. We anticipate this is partly because our assumption on the prior of individual locations does not necessarily hold.

\subsubsection{Effect of Noise on Privacy Loss} \label{sec:q3}
\begin{figure}[t]
         \centering
         \includegraphics[width=0.48\textwidth]{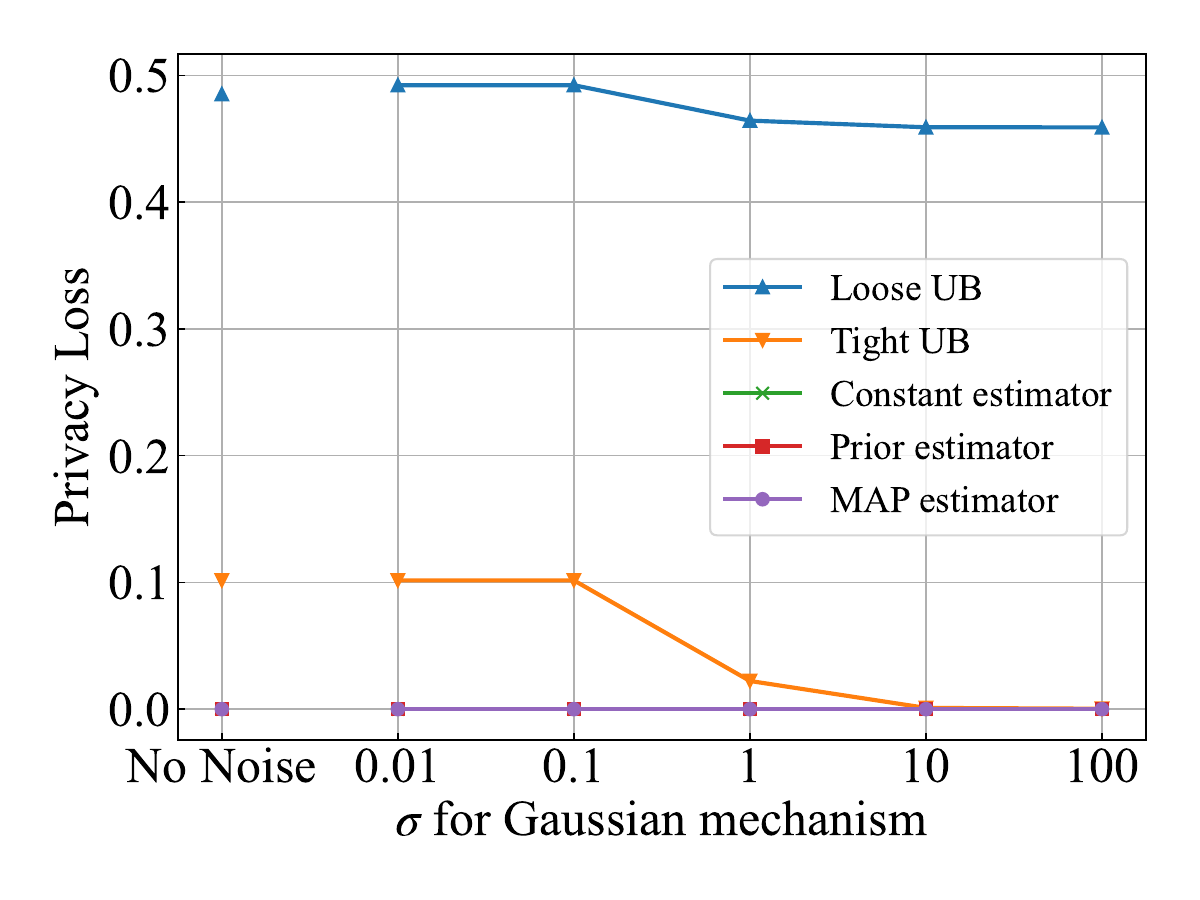}
         \caption{Privacy loss for the adversarial estimators and upper bounds with varying standard deviations of the additive Gaussian noise on the simulated data. We plot the estimates and upper bounds under no noise addition for reference. The privacy loss upper bounds decrease as the noise standard deviation becomes larger, but the privacy loss for the adversarial estimators remains constant.}
         \label{fig:exp-sigma-sim}
\end{figure}

\begin{figure}[t]
     \begin{subfigure}[b]{0.48\textwidth}
         \centering
         \includegraphics[width=\textwidth]{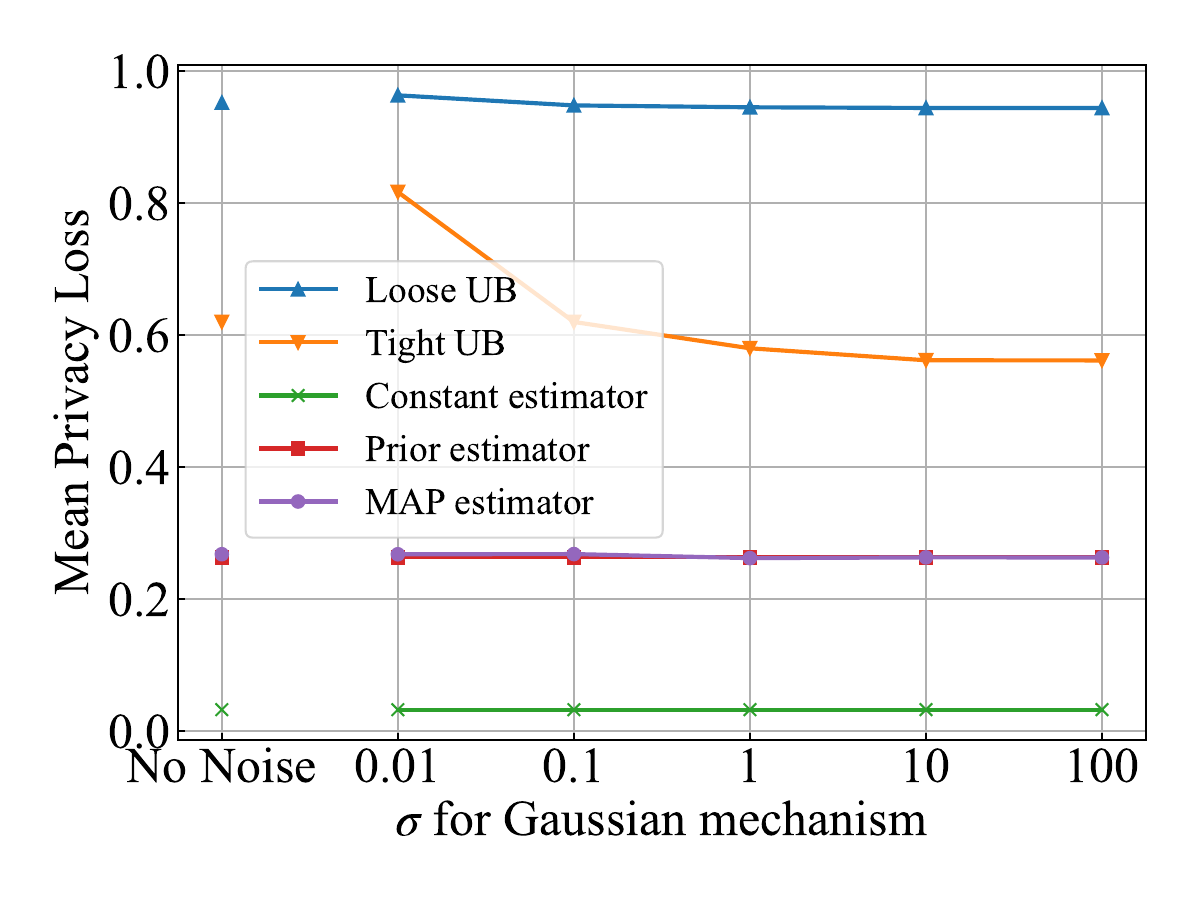}
         \caption{Foursquare}
     \end{subfigure}\hfill
     \begin{subfigure}[b]{0.48\textwidth}
         \centering
         \includegraphics[width=\textwidth]{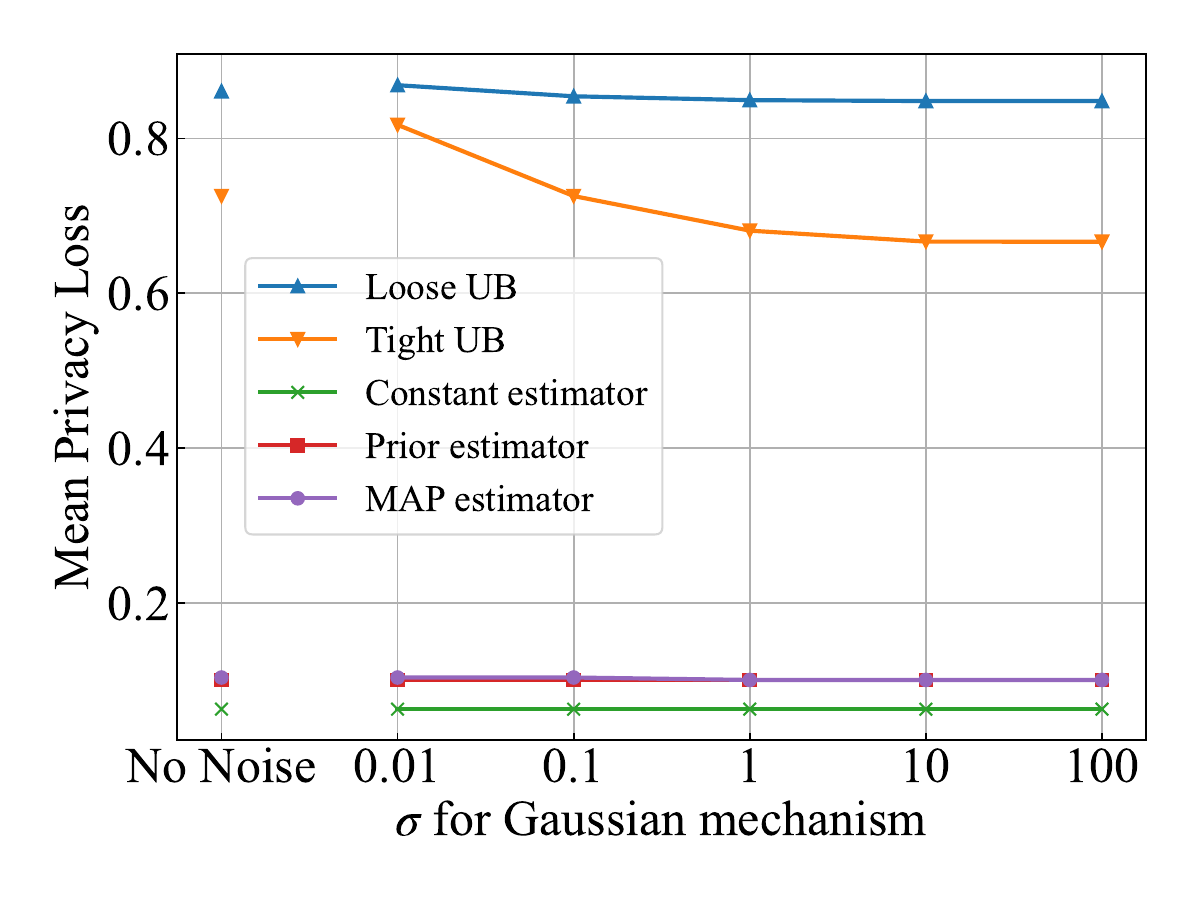}
         \caption{Gowalla}
     \end{subfigure}
    \caption{Mean privacy loss for the adversarial estimators and upper bounds over individuals with varying standard deviations of the additive Gaussian noise on the Foursquare and Gowalla datasets. We plot the estimates and upper bounds under no noise addition for reference. The mean privacy loss upper bounds decrease as the noise standard deviation becomes larger, but the mean privacy loss for the adversarial estimators remains constant.}
  \label{fig:exp-sigma-real}
\end{figure}
Figures~\ref{fig:exp-sigma-sim} and ~\ref{fig:exp-sigma-real} show how adding noise to the counts impacts the privacy loss.
We see in general that our privacy loss upper bounds get smaller as we add noise with a larger standard deviation $\sigma$.
The bounds are higher than the ones without any additive noise for some $\sigma$'s on real datasets. This is because the privacy loss upper bound for the noisy counts is not tight enough compared with the one for the raw counts. However, we expect the actual privacy loss bounds to be lower than the ones without any additive noise even for such $\sigma$'s.
Furthermore, we observe that adding noise does not change the privacy loss for the MAP estimator much, while it inevitably distorts the published counts. This suggests that if the adversary has enough information from the prior distribution, the noise in the published counts cannot make them forget the information they already know.
By increasing the noise standard variation, the DP loss decreases. However, as we see from the privacy loss for the MAP estimator, adding noise does not necessarily help reduce the privacy loss. This result implies that the DP loss might be overly conservative.

\section{Conclusion and Future Work}
This paper formalizes a distributed spatio-temporal counting problem and sets up an adversary who aims to reconstruct an individual's locations through the released counts. We then define a new privacy loss against the reconstruction attacks by the adversary, which better reflects our intuitions on privacy violations in this setting. Furthermore, we propose a privacy loss estimation algorithm against adversarial estimators and derive upper bounds on our privacy loss for any estimator. Our experimental results demonstrate that our privacy loss is more intuitive and fine-grained than DP in the distributed spatio-temporal setting and it behaves naturally with different parameters. 

One of our future directions is to provide a more thorough guideline for statistics release based on our privacy loss. Moreover, we currently assume that the adversary knows all individuals' exact locations except for the target's. We plan to relax the adversary further to incorporate the uncertainty of other people than the target to measure more realistic privacy loss against reconstruction attacks by an adversary who has more limited information.

\section*{Acknowledgments}
TK and KC would like to thank 
NSF under 1804829,
NSF under 2241100,
NSF under 2217058,
ARO MURI W911NF2110317, and
ONR under N00014-20-1-2334
for research support. 
Also, TK is supported in part by Funai Overseas Fellowship.

\bibliographystyle{plain}
\bibliography{ms}

\end{document}

\appendix
\section{Proof of Proposition 1}
\begin{proof}
We show by constructing a concrete algorithm.
Let $f(t,x_t) = \Pr[Y_t = y_t| X_t = x_t]\max_{x_1,\ldots, x_{t-1}} \prod_{u=1}^{t-1} \Pr[Y_u = y_u| X_u = x_u]\cdot \prod_{u=1}^{t} \Pr[X_u=x_u | X_{u-1}=x_{u-1}]$.
Note that $f(1,x_1) = \Pr[Y_1 = y_1| X_1 = x_1]\Pr[X_1=x_1]$.
Then, we have $\max_{x_T}f(T,x_T) = \max_{x_1,\ldots, x_T} \prod_{t=1}^{T} \Pr[Y_t = y_t| X_t = x_t]\cdot \prod_{t=1}^{T} \Pr[X_t=x_t | X_{t-1}=x_{t-1}]$.
For $t > 1$, we have the following:
\begin{align*}
    f(t,x_t) &= \Pr[Y_t = y_t| X_t = x_t]\max_{x_1,\ldots, x_{t-1}} \prod_{u=1}^{t-1} \Pr[Y_u = y_u| X_u = x_u]\cdot \prod_{u=1}^{t} \Pr[X_u=x_u | X_{u-1}=x_{u-1}]\\
    &= \Pr[Y_t = y_t| X_t = x_t] \max_{x_{t-1}}\Pr[X_t=x_t | X_{t-1}=x_{t-1}] \Pr[Y_{t-1} = y_{t-1}| X_{t-1} = x_{t-1}] \\
    &\quad \max_{x_1,\ldots, x_{t-2}} \prod_{u=1}^{t-2} \Pr[Y_u = y_u| X_u = x_u]\cdot \prod_{u=1}^{t-1} \Pr[X_u=x_u | X_{u-1}=x_{u-1}]\\
    &= \Pr[Y_t = y_t| X_t = x_t] \max_{x_{t-1}}\Pr[X_t=x_t | X_{t-1}=x_{t-1}] f(t-1,x_{t-1}).
\end{align*}
Note that $\Pr[Y_t = y_t| X_t = x_t]$ is defined by the mechanism.
Therefore, by sequentially computing $f(t,x_t)$ for $t$ from $t=1$ to $t=T$ and each $x_t$, we obtain $\max_{x_1,\ldots, x_T} \prod_{t=1}^{T} \Pr[Y_t = y_t| X_t = x_t]\cdot \prod_{t=1}^{T} \Pr[X_t=x_t | X_{t-1}=x_{t-1}]$. 
At the same time, we retain $h(t, x_t) = \argmax_{x_{t-1}}\Pr[X_t=x_t | X_{t-1}=x_{t-1}] f(t-1,x_{t-1})$ for $t>1$ and each $x_t$. 
To obtain the MAP estimate $\hat{X}^\mathrm{MAP}$, we carry out backtracing with $h$. More specifically, $\hat{X}^\mathrm{MAP}_T = \argmax_{x_T} f(T,x_T)$, and for $t<T$, $\hat{X}^\mathrm{MAP}_t = h(t+1, \hat{X}^\mathrm{MAP}_{t+1})$.
It takes $\mathcal{O}(TM^2)$ time and $\mathcal{O}(TM)$ space.
\end{proof}

\section{Proof of Proposition 2}
\begin{proof}
We show by constructing a concrete algorithm.
Let $f(t, x_t) = \max_{x_1,\ldots,x_{t-1}} \Pr[X_1=x_1,\ldots,X_t=x_t]$ for $t \geq 1$.
Note that $f(1,x_1) = \Pr[X_1=x_1]$. 
Then, we have $\max_{x} \Pr[X=x] = \max_{x_T} f(T,x_T)$.
Furthermore, for $t>1$, we have the following:
\begin{align*}
    f(t,x_t) &= \max_{x_1,\ldots,x_{t-1}} \Pr[X_1=x_1,\ldots,X_t=x_t]\\
    &=\max_{x_{t-1}} \Pr[X_t=x_t|X_{t-1}=x_{t-1}] \max_{x_1,\ldots,x_{t-2}} \Pr[X_1=x_1,\ldots,X_{t-1}=x_{t-1}]\\
    &=\max_{x_{t-1}} \Pr[X_t=x_t|X_{t-1}=x_{t-1}] f(t-1,x_{t-1}).
\end{align*}
Note that $\Pr[X_t=x_t|X_{t-1}=x_{t-1}]$ is the transition matrix.
Therefore, by sequentially computing $f(t,x_t)$ for $t$ from $t=1$ to $t=T$ and each $x_t$, we obtain $\max_{x} \Pr[X=x]$.
Simultaneously, we compute $h(t,x_t) = \argmax_{x_{t-1}}\Pr[X_t=x_t|X_{t-1}=x_{t-1}] f(t-1,x_{t-1})$ for $t > 1$ and each $x_t$.
To obtain $\hat{X}^\mathrm{prior}$, we carry out backtracing with $h$, i.e., $\hat{X}^\mathrm{prior}_T = \argmax_{x_T}f(T,x_T)$ and $\hat{X}^\mathrm{prior}_{t}=h(t+1,\hat{X}^\mathrm{prior}_{t+1})$ for $t < T$.
The total procedure is done with time $\mathcal{O}(TM^2)$ and space $\mathcal{O}(TM)$. 
\end{proof}

\section{Upper bound on the mutual information under the Gaussian mechanism}
We use the following upper bound on the mutual information when the Gaussian mechanism~\cite{dwork_algorithmic_2014} is used.
\begin{align*}
    &I(X;\mathrm{count}(D_1,c_1),\ldots,\mathrm{count}(D_T,c_T))\\
    &\leq \sum_{t=1}^{T}
    -p_t\log\left(p_t + (1-p_t)\exp\left(\frac{-1}{2\sigma^2}\right)\right) 
    - (1-p_t)\log\left((1-p_t) + p_t\exp\left(\frac{-1}{2\sigma^2}\right)\right),
\end{align*}
The proof, which is analogous to the one of Theorem 2 in \cite{guo_analyzing_2022}, is as follows.
\begin{proof}
    First, we have:
    \begin{align*}
        I(X;\mathrm{count}(D_1,c_1),\ldots,\mathrm{count}(D_T,c_T))
    \leq \sum_{t=1}^{T} I(X_t;\mathrm{count}(D_t,c_t)).
    \end{align*}
    From Section 3.1 in \cite{nielsen_w-mixtures_2021} and Theorem 2 in \cite{guo_analyzing_2022}, it holds that 
    \begin{align*}
        I(X_t;\mathrm{count}(D_t,c_t)) &\leq 
        -\sum_{m=1}^{M}\Pr[X_t=m]\log\left(\sum_{l=1}^{M}\Pr[X_t=l]\exp\left(-D_1(\mathcal{N}(\textbf{1}[m=c_t],\sigma^2),\mathcal{N}(\textbf{1}[l=c_t],\sigma^2)\right)\right) \\
        &=-p_t\log\left(p_t + (1-p_t)\exp\left(\frac{-1}{2\sigma^2}\right)\right) 
    - (1-p_t)\log\left((1-p_t) + p_t\exp\left(\frac{-1}{2\sigma^2}\right)\right),
    \end{align*}
    where $D_1$ is the KL divergence and $\textbf{1}[\cdot]$ is the indicator function.
    Combining the inequalities above yields the final result.
\end{proof}

\section{Proof of Theorem 1}

\begin{proof}    
We show by constructing a concrete algorithm. 
By changing the variables, we aim to obtain $f(m) = \max_{1\leq t_1,\ldots,t_m\leq T}\max_{x_1,\ldots,x_m} \Pr[X_{t_1}=x_1,\ldots,X_{t_m}=x_m]$ for $1\leq m \leq T$.
For $m=1$, because we assume a stationary MC, $f(1) = \max_{x_1}\Pr[X_1=x_1]$.
For $m>1$, by rearranging, we have:
\begin{align*}
    f(m) &= \max_{1\leq t_1,\ldots,t_m\leq T}\max_{x_1,\ldots,x_m} \Pr[X_{t_1}=x_1,\ldots,X_{t_m}=x_m]\\
    &=\max_{\Delta_1>0,\ldots,\Delta_{m-1}>0:\sum_{i=1}^{m-1}\Delta_i \leq T-1}\max_{x_1,\ldots,x_m} \Pr[X_{1}=x_1,\ldots,X_{1+\Delta_1+\cdots+\Delta_{m-1}}=x_m]\\
    &=\max_{m-1\leq k \leq T-1}\max_{\Delta_1,\ldots,\Delta_{m-1}:\sum_{i=1}^{m-1}\Delta_i=k}\max_{x_1,\ldots,x_m} \Pr[X_{1}=x_1,\ldots,X_{1+\Delta_1+\cdots+\Delta_{m-1}}=x_m]
\end{align*}
where the second equality follows due to a stationary MC.
Next, for $m>1$, $m-1\leq k\leq T-1$, and $x_m \in [M]$, we let
\begin{align*}
    g(m,k,x_m) = \max_{\Delta_1,\ldots,\Delta_{m-1}:\sum_{i=1}^{m-1}=k}\max_{x_1,\ldots,x_{m-1}} \Pr[X_{1}=x_1,\ldots,X_{1+\Delta_1+\cdots+\Delta_{m-1}}=x_m]
\end{align*}
so that $f(m) = \max_{m-1\leq k \leq T-1}\max_{x_m} g(m,k,x_m)$.
Note that $g(2,k,x_2)=\max_{x_1} \Pr[X_1=x_1,X_{1+k}=x_2]=\max_{x_1} \Pr[X_{1+k}=x_2|X_1=x_1]\Pr[X_1=x_1]$ is computed with the initial prior and $k$-th power of the transition matrix.
For $m>2$, we have the following:
\begin{align*}
    g(m,k,x_m) &= \max_{\Delta_1,\ldots,\Delta_{m-1}:\sum_{i=1}^{m-1}=k}\max_{x_1,\ldots,x_{m-1}} \Pr[X_{1}=x_1,\ldots,X_{1+\Delta_1+\cdots+\Delta_{m-1}}=x_m]\\
    &= \max_{1\leq \Delta_{m-1}\leq k-m+2}\max_{x_{m-1}} \max_{\Delta_1,\ldots,\Delta_{m-2}:\sum_{i=1}^{m-2}=k-\Delta_{m-1}}\max_{x_1,\ldots,x_{m-2}}\\
    &\quad \Pr[X_{1+\Delta_1+\cdots+\Delta_{m-1}} = x_m|X_{1+\Delta_1+\cdots+\Delta_{m-2}} = x_{m-1}] \Pr[X_{1}=x_1,\ldots,X_{1+\Delta_1+\cdots+\Delta_{m-2}}=x_{m-1}]\\
    &= \max_{1\leq \Delta_{m-1}\leq k-m+2}\max_{x_{m-1}} \Pr[X_{1+\Delta_{m-1}} = x_m|X_{1} = x_{m-1}]\\
    &\quad \cdot \max_{\Delta_1,\ldots,\Delta_{m-2}:\sum_{i=1}^{m-2}=k-\Delta_{m-1}}\max_{x_1,\ldots,x_{m-2}} \Pr[X_{1}=x_1,\ldots,X_{1+\Delta_1+\cdots+\Delta_{m-2}}=x_{m-1}]\\
    &= \max_{1\leq \Delta_{m-1}\leq k-m+2}\max_{x_{m-1}} \Pr[X_{1+\Delta_{m-1}} = x_m|X_{1} = x_{m-1}] g(m-1,k-\Delta_{m-1},x_{m-1}).
\end{align*}
Therefore, by sequentially computing $g(m,k,x_m)$ for $m$ from $m=1$ to $m=T$ and each $k$ and $x_m$, we obtain $f(m)$ and thus, $\max_{\hat{x}} \Pr[X\in\{x:d(x,\hat{x}) \leq s\}]$.
It takes $\mathcal{O}(T^3M^2)$ time and (additional) $\mathcal{O}(TM)$ space given an access to first to the $T-1$-th power of the transition matrix. 
The powers of the transition matrix requires $\mathcal{O}(T\log TM^3)$ time and $\mathcal{O}(TM^2)$ space; thus, in total it takes $\mathcal{O}(T^3M^2 + T\log TM^3)$ time and $\mathcal{O}(TM^2)$ space.
\end{proof}

\bibliographystyle{plain}
\bibliography{supplement}